\documentclass[twocolumn,superscriptaddress,showpacs,amsmath,amssymb,aps,prd]{revtex4}
\usepackage{graphicx}
\usepackage{bm}
\usepackage{color}
\usepackage{multirow}
\usepackage{slashbox}

\def\bh#1{black hole#1
  (BH#1)\gdef\bh{BH}}
\def\bbh#1{binary black hole#1
  (BBH#1)\gdef\bbh{BBH}}
\def\gw#1{gravitational wave#1
  (GW#1)\gdef\gw{GW}}
\def\nr#1{numerical relativity#1
  (NR#1)\gdef\nr{NR}}
  \def\snr#1{signal-to-noise#1
  (SNR#1)\gdef\snr{SNR}}
    \def\gw#1{gravitational wave#1
  (GW#1)\gdef\gw{GW}}

\begin{document}

\title{Template Mode Hierarchies for Binary Black Hole Mergers}

\author{James Healy}
\affiliation{Center for Relativistic Astrophysics and
School of Physics\\
Georgia Institute of Technology, Atlanta, GA 30332}
\author{Pablo Laguna}
\affiliation{Center for Relativistic Astrophysics and
School of Physics\\
Georgia Institute of Technology, Atlanta, GA 30332}
\author{Larne Pekowsky}
\affiliation{Center for Relativistic Astrophysics and
School of Physics\\
Georgia Institute of Technology, Atlanta, GA 30332}
\author{Deirdre Shoemaker}
\affiliation{Center for Relativistic Astrophysics and
School of Physics\\
Georgia Institute of Technology, Atlanta, GA 30332}

\begin{abstract} 
 Matched filtering is a popular data analysis framework
used to search for gravitational wave signals emitted by compact object binaries.  
The templates used in matched filtering searches are constructed predominantly from the quadrupolar mode since 
this is the energetically most dominant mode. However, for highly precessing binaries or binaries with moderately large mass ratios,
quadrupole templates lose sensitivity for a significant fraction of source orientations. 
We investigate how the inclusion of higher modes in the templates  alleviates this loss of sensitivity and thus increases the prospects for detecting gravitational waves.
Specifically, we use numerical relativity waveforms from the late inspiral and coalescence of binary black holes
to identify mode hierarchies from which one can construct templates that cover the entire sky of binary orientations. The ordering in these hierarchies 
depends on the characteristics of the binary system and the mode strengths. The proposed hierarchies could assist deciding
which modes to add to templates banks according to their ability for maximizing sky-coverage.
\end{abstract}

\maketitle

\section{Introduction}

Binary systems with \bh{} components will provide us with one of the strongest sources of 
\gw{s} soon to be detected by ground-based interferometers~\cite{lrr-2011-5}.  
\gw{} observations are accompanied  by dazzling  engineering, theoretical, and data analysis challenges. One challenge that requires the
 close cooperation between source model and data analysis researchers is
digging out  the \gw{} signals from the noisy data. 
For compact object binaries involving neutron stars and \bh{s},
one of the most popular frameworks to analyze the data is \emph{matched filtering}, which in the broad sense consists of correlating 
a known signal or \emph{template} with the data in hopes of identifying  the presence of the template in the data.
Not surprisingly, the success of matched filtering depends mainly on the ``quality'' of the template, in other words, the degree to which 
the template captures the correct physics and matches the characteristics of the source. 
Even with the belief that we possess the correct theory of gravity, Einstein's theory of general relativity, 
constructing templates is a laborious and arduous task. A family of \bbh{} templates needs to sample 
a 15-dimensional parameter space (\bh{} masses and spins, binary eccentricity, orientation vector, sky position and distance).
Furthermore,  if the binary merges in the sweet spot of the detector,
building templates requires \nr{} simulations that capture dynamically strong gravity, thus increasing the computational cost per template significantly.

\begin{figure}[tb]
\centering
\vbox{
\includegraphics[width=.95\linewidth]{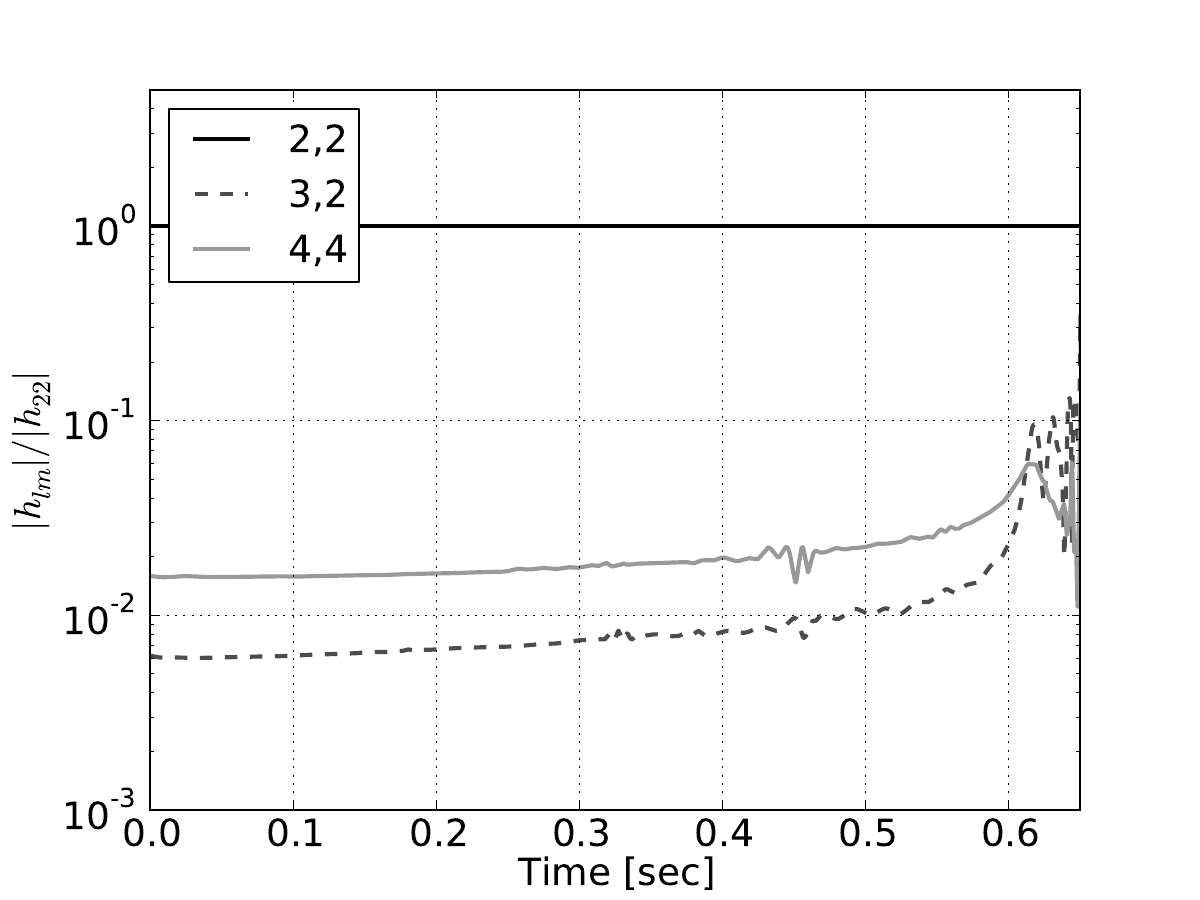}\\
\includegraphics[width=.95\linewidth]{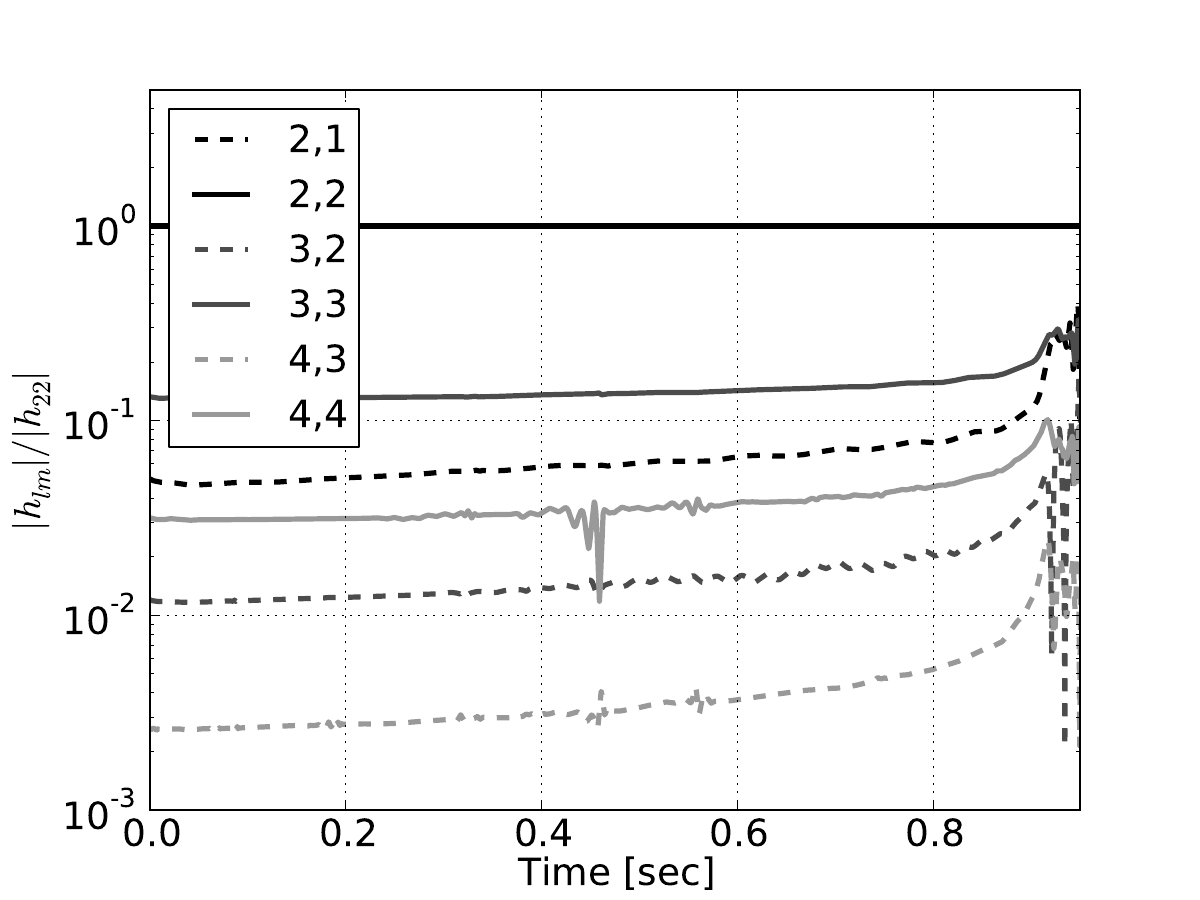}
}
\caption{Strain mode amplitude ratios relative to the (2,2) mode non-spinning \bbh{s}.
Top panel shows the case of an equal mass binary, and the bottom panel
that of a 1:4 mass ratio system.  Modes not included have relative
amplitudes less than $10^{-3}$.  In both figures the systems have been
scaled to total mass $M = 100 M_\odot$.
}
\label{fig:modeAmplitudes}
\end{figure}

Without accounting for effects from the orientation of the source, the gravitational radiation from a binary system is dominated by its quadrupolar (2,2) mode component.
It should not be surprising then that current data analysis pipelines are predominantly using this mode as a template.
An example of the dominance of the (2,2) mode is depicted in Figure~\ref{fig:modeAmplitudes},
where we show the strain amplitude ratio of higher modes relative to the (2,2) mode.
The top panel shows the case of an equal mass \bbh{} and the bottom one with a 1:4 mass ratio. It is evident in both cases the dominance of the (2,2) mode. However, 
for the unequal mass case, the second strongest mode, the (3,3) mode, is more than 10\% of the (2,2), reaching above 20\% near merger.
On the other hand, for the equal mass case, the second strongest mode is the (4,4) mode, and it is only a few percent of the (2,2) in strength. 
Similar situations occur for precessing (i.e. spinning \bh{s}) binaries. 

\begin{figure}[tb]
\centering
\vbox{
\includegraphics[width=.95\linewidth]{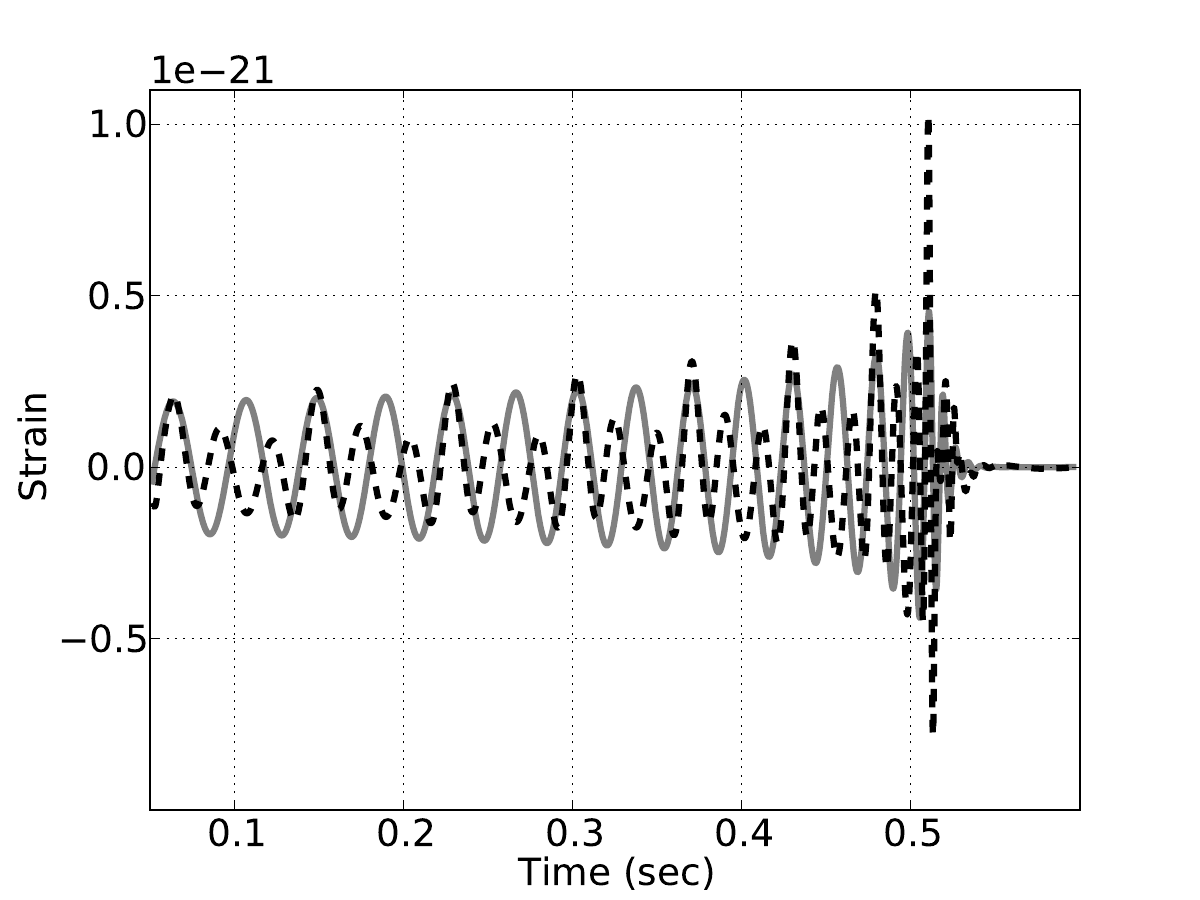}\\
}
\caption{Strain of a 1:10 mass ratio system with non-spinning \bh{s},
scaled to $M = 100 M_\odot$. The (2,2) mode is shown as solid grey line and a waveform including all modes  $2 \le \ell \le 5$ with a dashed balck line.
The waveforms are depicted as seen from an orientation $\theta=26^\circ,
\phi=45^\circ$ in coordinated system centered at the source.  The
(2,2) line has been multiplied by 15 in order to better highlight the
difference in shape between the (2,2) mode and the full waveform. The
overlap between these two signals is 0.967, which is below the desired
overlap of 0.97.}
\label{fig:recomposedQ10}
\end{figure}

The relative strength of a mode can also be modified depending on the orientation of the source. 
For instance, in a coordinate system centered at the source,
the (2,2) mode has a dependence $\propto (1+ \cos\theta)^2$, with  $\theta$ the polar angle. Thus, as the line-of-sight moves away from the north pole, 
the amplitude of the (2,2) mode will decrease relative to a template with higher mode content.
Figure~\ref{fig:recomposedQ10} illustrates this effect for a binary system of non-spinning \bh{s} with 1:10 mass ratio. 
The figure shows waveforms as observed along a line-of-sight with orientation $\theta=26^\circ$ and $\phi=45^\circ$.
The dashed black line depicts a {\it full} waveform that includes all modes between $2\le \ell \le 5$; the solid grey line represents the (2,2) mode waveform, 
which has been enhanced by a factor of 15 for clarity. 
Notice that, in addition to the decrease in amplitude due to the orientation, the {\it shape} of the (2,2) mode differs significantly
from the full waveform. Furthermore, the overlap between the (2,2) mode and the full waveform is 0.967, a clear indication of the loss of sensitivity by the (2,2) mode.

Recently, we carried out a study addressing the loss of sensitivity of quadrupolar templates~\cite{2012arXiv1210.1891P}. 
We found that for precessing binaries, overlaps could drop below 0.97 for up to 65\% of the sky at the source location.
The present study takes the next step.  
We investigate how the limitations in sky coverage of quadrupolar templates  can be alleviated by the inclusion of higher modes.
Since adding higher modes to templates has the potential of affecting the efficiency of template bank searches, it is important to 
also investigate the payoff gained by adding a given higher mode. When adding more than one higher mode, the issue of ordering plays also a role
in evaluating efficiency. Therefore, the goal of the present study is not only to identify the modes for full sky coverage, but also to find
the most efficient ordering of modes (i.e. hierarchy of modes) to
build a template. 

The difficulty of constructing a
multi-mode search and the computation cost of such a search are
likely to be insensitive to the number of modes once the first mode is added.  
However, this is not true in building templates,
which must be done on a mode-by-mode basis regardless of whether this
is done by post-Newtonian methods or fitting models against waveforms
obtained from numerical relativity.
Hierarchies such as the ones we are proposing could help deciding
which modes to add according to their ability to maximize
sky-coverage, and by implication which modes warrant the most effort
in development and in what order this effort should be prioritized.

Our study also shows that building hierarchies of higher modes based solely on their 
magnitudes may not be optimal because it ignores the contributions from the orientation of the source (i.e. angular dependence of the 
spin-weighted spherical harmonics). In other words,
it is possible for a mode with relatively small amplitude to become
more prominent due to the spherical harmonic involved that allows
to fill in a gap in coverage more effectively.  Our study will demonstrate how this
occurs in practice.

 Our study assumes that the detector is \emph{optimally
oriented}, and we focus on line-of-sight, source orientation effects.
Furthermore, we focus on \bbh{s} with total mass $M = 100 M_\odot$. For these relatively massive binary systems,  
interferometers such as LIGO only ``see'' the late inspiral and merger of the binary. 
Therefore,  templates in this study are obtained entirely from \nr{} simulations. For less massive binaries, the early inspiral becomes more important. 
The templates in those cases would have to be obtained from hybrid waveforms that stitch \nr{} and post-Newtonian strains~\cite{Hannam:2010ky,Boyle:2011dy}.

Our study finds that template banks for \gw{} detection based only on the (2,2) mode are probably sufficient for comparable mass, low precession \bbh{s}. 
At the same time, our work demonstrates that the effectiveness of match filtering is severely impaired for highly precessing or moderately large unequal mass \bh{} systems if 
templates banks are based entirely on the quadrupolar mode.

\section{Matched Filtering and Detection Reach}
\label{sec:matchedfilter}

A \gw{} signal $s$ impinging an interferometer has the following structure
\begin{eqnarray}
\label{eq:recievedSignal}
s(t;\bar\theta,\bar\phi,\psi,{\vec\xi})
&=& F_+(\bar\theta,\bar\phi,\psi)\, h_+(t;\vec{\xi}) \nonumber \\
    &+& F_\times(\bar\theta,\bar\phi,\psi)\, h_\times(t;\vec{\xi})\,,
\end{eqnarray}
where $h_+$ and $h_\times$ denote the two polarization strains of general relativity.  $h_+$ and $h_\times$ depend on a parameter vector $\vec{\xi}$. Some of the components in $\vec{\xi}$  are parameters
intrinsic to the system, such as the binary eccentricity, \bh{} masses and spins; others are external parameters like the binary orientation 
and its distance to the detector. In addition, the signal $s$ depends on the characteristics of the detector through  
 the response or {\it antenna pattern} functions $F_+$ and $F_\times$~\cite{PhysRevD.63.042003}. 
 For an interferometer with arms oriented along the $x$ and $y$ axes, these functions read
\begin{eqnarray}
\label{eq:AntPatternFp}
F_+ &=& -\frac{1}{2} (1+\cos^2 \bar\theta) \cos 2\bar\phi \cos 2\psi \nonumber\\
&-&\cos\bar\theta \sin 2\bar \phi \sin 2\psi \\
\label{eq:AntPatternFm}
F_\times &=& \frac{1}{2} (1+\cos^2 \bar\theta) \cos 2\bar\phi \sin 2\psi  \nonumber\\
&-& \cos\bar\theta \sin 2\bar \phi \cos 2\psi \,,
\end{eqnarray}
where the angles $(\bar\theta, \bar\phi)$ are 
the location of the binary system in the sky of the detector and $\psi$ the polarization angle.
A detector is optimally oriented if a signal arrives with $\bar\theta = 0$ or $\pi$ and $\bar\phi=0$. On the other hand, the detector is blind for signals with
$\bar\theta=\pi/2$ and $\bar\phi$ any of $\{\pm \pi/4, \pm 3\pi/4\}$, that is,
arriving from directions in the plane of the interferometer and
directly between the arms. 
With a network of interferometers, the issue of  detector blindness can be alleviated~\cite{2011CQGra..28l5023S,2011arXiv1112.3092K}. For this reason, our focus will be
on ``intrinsic'' source orientation effects, which depend only on the direction of propagation of the \gw{.} That is, the signals arriving at the detector will be 
\begin{eqnarray}
\label{eq:recievedSignal2}
s(t;\theta,\phi,r,{\vec\xi}_b)
&=& h_+(t;\theta,\phi,r,{\vec\xi}_b) \,\cos 2\psi \nonumber\\
&+& h_\times(t;\theta,\phi,r,{\vec\xi}_b)\, \sin 2\psi\,.
\end{eqnarray}
In Eq.~(\ref{eq:recievedSignal2}), the angles $(\theta, \phi)$ give the direction of propagation of the \gw{} pointing to the detector from a reference system located at the source, $r$ is the distance between the source and the 
detector, and 
the vector $\vec{\xi}_b$ denote the parameters intrinsic to the binary (spins, masses and eccentricity).
For the present study, the reference system used at the source is the one used to carry out the \nr{} simulations.  That is, the reference system
has its origin at the center of mass of the 
binary, and has its $z$-axis aligned with the orbital angular momentum at the start of the simulation. 

Given the Fourier transforms $\tilde A$ and $\tilde B$ of two real time-dependent signals or waveforms,  the inner product with
respect to the noise spectrum density $S_n(f)$ of the detector is given by
\begin{equation}
\label{eq:InnerProduct}
\langle A|B \rangle
 = 4\, \mathrm{Re} \int_{0}^\infty df\,
   \frac
     {\tilde{A}(f) \tilde{B}^\star(f)} 
     {S_n(f)} \,.
\end{equation}
With this inner product, the \snr{} of  a signal $s$ and a template $u$ is given by~\cite{Maggiore2008}
\begin{equation}
\label{eq:Snr}
\rho(s,u)
= \max_{t_0,\psi_0}
\frac{\langle s|u\rangle}
     {\sqrt{\langle u|u\rangle}}\,.
\end{equation}
The maximization over $t_0$ enters because the signal arrives at an unknown time.  Thus, one needs to ``slide'' the template relative to the signal, which in Fourier space implies replacing $\tilde{u}(f)$ by $\tilde{u}(f)\exp(-2 \pi ift_0)$ in Eq.~(\ref{eq:Snr}).
In principle, there is also an unknown phase difference $\psi_0$
between the signal and the template related to the polarization angle $\psi$.  This introduces an additional factor of $\exp(2
\pi i \psi_0)$. If $u=s$ in Eq~(\ref{eq:Snr}), one gets the so-called  \emph{optimal} \snr{} $\rho(s,s) = \sqrt{(s|s)}$.

The {\it match} or {\it overlap} $\mu$ between the signal $s$ and a template $u$ is given by
\begin{equation}
\label{eq:overlap}
\mu(s,u)
 = \max_{t_0,\psi_0} \frac{\langle s|u \rangle}
{\sqrt{ \langle s|s\rangle\, \langle u|u\rangle}}\,.
\end{equation}
Thus, $\rho(s,u) = \mu(s,u)\,\sqrt{\langle s|s \rangle} = \mu(s,u)\,\rho( s,s)$. In some instances, it is more convenient to work with \emph{mismatches}, which are defined as $\epsilon \equiv 1-\mu$. Notice that the $1/r$ scaling of the signal $s$ implies that $\rho \propto 1/r$. However, because of the normalization factors,  
$\mu$ and $\epsilon$ are independent of $r$. Furthermore, the distance scale $r$ used to construct a template does not play any role because 
of the normalization factor $\sqrt{\langle u|u\rangle}$ in Eqs.~(\ref{eq:Snr}) and (\ref{eq:overlap}). 

Consider two signals $s_1$ and $s_2$ from two identical \gw{} sources, $s_1$ located at distances $r_1$ and $s_2$ at $r_2$. Then,
\begin{equation}
\label{eq:rho1}
\frac{\rho(s_1,u)}{\rho(s_2,u)} = \frac{\mu(s_1,u)}{\mu(s_2,u)}\frac{\rho(s_1,s_1)}{\rho(s_2,s_2)}= \frac{\rho(s_1,s_1)}{\rho(s_2,s_2)}\,,
\end{equation}
where the last equality follows because the sources are identical, and thus $\mu(s_1,u) = \mu(s_2,u)$. On the other hand,
\begin{eqnarray}
\label{eq:snrprop}
\rho(s_2,s_2) &=&\frac{1}{r_2} \rho(r_2\,s_2,r_2\,s_2) \nonumber \\
&=& \frac{1}{r_2} \rho(r_1\,s_1,r_1\,s_1) = \frac{r_1}{r_2} \rho(s_1,s_1)\,,
\end{eqnarray}
where we have used that the optimal \snr{} of a signal times its distance, $r\,s$, is independent of $r$. Therefore, we can rewrite Eq.~(\ref{eq:rho1}) as
\begin{equation}
\label{eq:ratio}
 \frac{\rho(s_1,u)}{\rho(s_2,u)} = \frac{\rho(s_1,s_1)}{\rho(s_2,s_2)}  = \frac{r_2}{r_1} \,.
\end{equation}
It is wrong to conclude from Eq.~(\ref{eq:ratio}) that the ratio $r_2/r_1$ depends on the template $u$ because of  $\rho(s_1,u)/\rho(s_2,u)$. 
The dependence on $u$ is eliminated because we are considering the same source, just located at different distances. 

The \snr{} property (\ref{eq:snrprop}) allow one to define for a signal $s$ a \emph{horizon} distance $R(s)$ as follows: \begin{equation}
\label{eq:distance}
R(s) = \frac{\rho(\hat s,\hat s) }{\rho(s,s)}
\end{equation}
where $\rho(\hat s, \hat s)$ is the optimal \snr{} of the ``signal" $\hat s \equiv r\,s$. $R(s)$ should be interpreted as providing the maximum distance that an interferometer is able
 to detect a signal $s$ given an optimal \snr{} threshold.  For instance, during LIGO's science run S6, the horizon distance of a compact binary coalescence with optimal \snr{} threshold of $\rho(s,s) =  8$~Mpc$^{-1}$  was estimated to be $R(s)\approx  40$~Mpc using as 
 a model for the signal a post-Newtonian strain in the stationary phase approximation~\cite{2012arXiv1203.2674T}. Therefore from Eq.~(\ref{eq:distance}), $\rho(\hat s,\hat s) \approx 320$.

For the present study, we are interested in
investigating the effect that the choice of a template $u$ has on the ability of an interferometer to detect a signal $s$. Therefore, instead of the \emph{horizon} distance $R(s)$, which is independent of the template, we need a  \emph{template reach} distance ${\cal R}(s,u)$. Since the mismatch $\mu$ characterizes the ``proximity'' of a template to a signal, 
we define the template reach distance as
\begin{equation}
\label{eq:ru}
{\cal R}(s,u) = \mu(s,u)\,R(s)\,.
\end{equation}
Given that $\mu(s, u) = \mu(\hat s, u)$, we can rewrite Eq.~(\ref{eq:ru}) in a form similar to that of Eq.~(\ref{eq:distance}):
\begin{equation}
\label{eq:ru2}
{\cal R}(s,u) =  \frac{\rho(\hat s, u) }{\rho(s,s)}\,.
\end{equation}
From Eqs.~(\ref{eq:ru}) and (\ref{eq:ru2}), it evident that if the template exactly matches the signal, i.e. $u = s$, then the horizon distance $R(s)$ and the 
template reach distance are equal. On the other hand, if  $u \ne s$ then ${\cal R}(s,u) < R(s)$. 

The relative change in reach of a template $u$ can be estimated from
\begin{equation}
\frac{\delta R}{R} = \frac{R(s) - {\cal R}(s,u)}{R(s)} = 1-\mu(s,u) = \epsilon(s,u)\,,
\end{equation}
which with the help of Eq.~(\ref{eq:ru2})  takes the form
\begin{equation}
\frac{\delta R}{R} = 1-\mu(s,u) = \epsilon(s,u)\,.
\end{equation}
In other words,  the relative loss in distance to which signals can be seen
above a given \snr{} threshold is proportional to the mismatch. 
For small values of $\epsilon$, the relative loss of distance is also proportional to 
the relative loss of volume since
\begin{equation}
\label{eq:volume}
\frac{\delta V}{V} \approx  \frac{\delta (R^3)}{R^3} \approx \frac{3\,\delta R}{R}  \approx 3\,\epsilon\,.
\end{equation}
For this reason, we will use $\epsilon$ as the basis in our investigation of the impact of higher modes.  

\section{Binary Black Hole Waveforms}
\label{sec:waveforms}

As mentioned in the introduction, the \bbh{s} we consider have total mass $M = 100\,M_\odot$. For these binaries, interferometers such as advanced LIGO  are most sensitive to \gw{} 
frequencies emitted by the binary during the late inspiral and merger; and, therefore, the templates and signals needed in our study are those constructed entirely from \nr{} simulations. 
The simulations were produced with the \texttt{Maya} code~\cite{Haas:2012bk,Healy:2011ef,Bode:2011xz,Bode:2011tq,Bode:2009mt,Healy:2009zm} of the \nr{} group at Georgia Tech.
\texttt{Maya} uses the \texttt{Einstein Toolkit} \cite{et-web}
which is based on the \texttt{CACTUS} \cite{cactus-web} infrastructure
and \texttt{CARPET} \cite{Schnetter-etal-03b} mesh refinement. Evolution thorns were generated with the
\texttt{Kranc} \cite{Husa:2004ip} code generator.

\begin{table}
   \begin{tabular}{|r|c|c|c|c|c|c|c|c|c|}
\hline
ID & $q$  & $\chi_1$ & $\chi_2$ & $\zeta_1$ & $\zeta_2$ & $d/M$ & $M/\Delta$ & $\tau_{2,2}$ & $\tau_{2,*}$\\
\hline
Q01 & 1.0 & 0.0& 0.0& 0& 0& 11 & 200& 1.000& 1.000\\
Q02 & 1.5 & 0.0& 0.0& 0& 0& 11 & 200& 1.000& 1.000\\
Q03 & 2.0 & 0.0& 0.0& 0& 0& 11 & 200& 0.634& 0.934\\
Q04 & 2.5 & 0.0& 0.0& 0& 0& 11 & 200& 0.356& 0.440\\
Q05 & 3.0 & 0.0& 0.0& 0& 0& 11 & 200& 0.258& 0.299\\
Q06 & 4.0 & 0.0& 0.0& 0& 0& 10 & 240& 0.177& 0.197\\
Q07 & 5.0 & 0.0& 0.0& 0& 0& 10 & 240& 0.144& 0.158\\
Q08 & 6.0 & 0.0& 0.0& 0& 0& 10 & 280& 0.123& 0.134\\
Q09 & 7.0 & 0.0& 0.0& 0& 0& 10 & 320& 0.111& 0.119\\
Q10 &10.0 & 0.0& 0.0& 0& 0& 8.4& 400& 0.097& 0.103\\
\hline
S01 & 1.5 & 0.2& 0.2& 0& 0& 11 & 200& 1.000& 1.000\\
S02 & 1.5 & 0.4& 0.4& 0& 0& 11 & 200& 1.000& 1.000\\
S03 & 4.0 & 0.2& 0.0& 0& 0& 10 & 240& 0.183& 0.198\\
S04 & 4.0 & 0.4& 0.0& 0& 0& 10 & 240& 0.191& 0.200\\
S05 & 4.0 & 0.6& 0.0& 0& 0& 10 & 240& 0.196& 0.201\\
S06 & 5.0 & 0.2& 0.0& 0& 0& 10 & 240& 0.149& 0.160\\
S07 & 5.0 & 0.4& 0.0& 0& 0& 10 & 240& 0.154& 0.161\\
S08 & 5.0 & 0.6& 0.0& 0& 0& 10 & 240& 0.160& 0.164\\
S09 & 6.0 & 0.2& 0.0& 0& 0& 10 & 280& 0.127& 0.136\\
S10 & 6.0 & 0.4& 0.0& 0& 0& 10 & 280& 0.134& 0.139\\
S11 & 6.0 & 0.6& 0.0& 0& 0& 10 & 280& 0.139& 0.142\\
\hline
P01 & 1.5 & 0.6& 0.6&  45&   0& 10 & 120& 1.000& 1.000\\
P02 & 1.5 & 0.6& 0.6&  60&   0& 10 & 120& 1.000& 1.000\\
P03 & 1.5 & 0.6& 0.6&  90&   0& 10 & 120& 0.779& 1.000\\
P04 & 2.0 & 0.6& 0.6&  45&   0& 10 & 120& 0.554& 0.982\\
P05 & 2.0 & 0.6& 0.6&  60&   0& 10 & 120& 0.375& 0.981\\
P06 & 2.0 & 0.6& 0.6&  90&   0& 10 & 120& 0.222& 0.933\\
P07 & 2.0 & 0.6& 0.6& 135&   0& 10 & 120& 0.370& 0.963\\
P08 & 2.0 & 0.6& 0.6& 270&   0& 10 & 120& 0.327& 0.933\\
P09 & 2.5 & 0.4& 0.4&  45&   0& 10 & 120& 0.312& 0.451\\
P10 & 2.5 & 0.4& 0.4&  60&   0& 10 & 120& 0.291& 0.456\\
P11 & 2.5 & 0.4& 0.4&  90&   0& 10 & 120& 0.208& 0.471\\
P12 & 2.5 & 0.6& 0.6&  45&   0& 10 & 120& 0.258& 0.465\\
P13 & 2.5 & 0.6& 0.6&  60&   0& 10 & 120& 0.213& 0.485\\
P14 & 2.5 & 0.6& 0.6&  90&   0& 10 & 120& 0.162& 0.510\\
P15 & 4.0 & 0.6& 0.6&  45&   0& 10 & 120& 0.099& 0.218\\
P16 & 4.0 & 0.6& 0.6&  60&   0& 10 & 120& 0.057& 0.224\\
P17 & 4.0 & 0.6& 0.6&  90&   0& 10 & 120& 0.019& 0.247\\
P18 & 4.0 & 0.6& 0.6&   0& 270& 10 & 140& 0.193& 0.204\\
P19 & 4.0 & 0.6& 0.6&  90& 270& 10 & 140& 0.001& 0.242\\
P20 & 4.0 & 0.6& 0.6& 150& 270& 10 & 140& 0.155& 0.225\\
P21 & 4.0 & 0.6& 0.6& 180& 270& 10 & 140& 0.188& 0.234\\
P22 & 4.0 & 0.6& 0.6& 210& 270& 10 & 140& 0.130& 0.218\\
P23 & 4.0 & 0.6& 0.6& 270& 270& 10 & 140& 0.026& 0.244\\
\hline
\end{tabular}

\caption{ Characteristics of the \bbh{} systems used in this study. $q=m_1/m_2$ denotes the \bh{} mass ratio, 
$\chi_{1,2}$ the dimensionless spin parameter, $\zeta_{1,2}$ the angle between the \bh{} spin 
and the $z$-axis in the $xz$-plane, $d/M$ the initial binary separation, and $\Delta/M$ the grid 
spacing at the finest mesh. The last two columns give the covering factor $\tau_{2,2}$ when 
only the (2,2) mode is used as the template and the covering factor $\tau_{2,*}$ when the template 
includes all the $| m | \le \ell = 2$ modes.}
\label{tab:NR}
\end{table}

We considered a variety of \bbh{} configurations in quasi-circular orbits. Table~\ref{tab:NR} gives the characteristics of each binary: \bh{} mass ratio $q = m_1/m_2$, dimensionless spin parameters $\chi$,  angle $\zeta$ between the \bh{} spin and the $z$-axis in the $xz$-plane, initial separations $d/M$, and grid spacing $\Delta/M$  in the finest refinement level, and the covering factors $\tau_{2,2}$ when only the (2,2) mode is used as the template and  $\tau_{2,*}$ which includes all the $| m | \le \ell = 2$ modes in the template 
(where the covering factor $\tau$ is defined in the next section). 
The configurations are classified into three series. Series Q consists of \bbh{} with non-spinning \bh{s}. For the series S, at least one of the \bh{s} is spinning and aligned with the orbital angular momentum; thus, there is no precession. Finally, the series P consists of precessing binary systems. 

The templates and signals were built  from strains $h_+$ and $h_\times$ obtained from the Weyl Scalar $\Psi_4$, one of the
outputs of the simulations. Specifically, we decompose the Weyl Scalar $\Psi_4$  produced during the simulation into spin-weighted spherical harmonics as
\begin{equation}
\label{eq:Psi4Decomposition}
r M \Psi_4(t;\theta,\phi) = \sum_{l,m}C_{\ell m}(t)\, {}_{-2} Y_{\ell m}(\theta,\phi) \,,
\end{equation}
where the angles $\theta$ and $\phi$ are relative to a coordinate system with origin at the center-of-mass of the binary and with a $z$-axis aligned with its
orbital angular momentum at the beginning of the simulation.
Then, we solve  $\Psi_4 = \ddot h_+ - i\,\ddot h_\times = \ddot h^\star$
to get the strains
\begin{equation}
\label{eq:hlm}
\frac{r}{M} h(t;\theta,\phi) = \sum_{l,m}H_{\ell m}(t)\, {}_{-2} Y^\star_{\ell m}(\theta,\phi)
\end{equation}

For our signals $s$, we calculate $h_+$ containing all the $( \ell,m )$ modes with $2 \le \ell \le 5$. 
The templates  $u$ are constructed  from different subsets among all the 32~$( \ell,m )$ modes for the signals. Explicitly, a template $u_n$ containing $n\le 32$ modes is given by
\begin{equation}
\label{eq:templates}
u_n = \sum_{i=1}^n w_i\,
\end{equation}
with $w_i$ denoting a mode in the sense of Eqs.~(\ref{eq:hlm}), i.e.,  $w_i = H_{\ell m}(t)\, {}_{-2} Y^\star_{\ell m}(\theta,\phi) M/r$ in which the index $i$ collectively labels $( \ell,m )$.
To avoid the integration to strain, and the errors associated with it, 
we calculate the strain in the frequency domain for the matches, that is, 
$\tilde{h} = \tilde{\Psi}_4^\star / (-4\pi^2f^2)\,.$ 

\section{Covering Factors and Mode Hierarchies}
\label{sec:greedy}

As mentioned in Sec.~\ref{sec:matchedfilter}, the focus of our analysis is the mismatch $\epsilon$ of a signal $s(\vec \xi_s)$ with a template $u(\vec\xi_u)$,
where the parameter vectors $\vec\xi_s$ and $\vec\xi_u$ include both,
intrinsic physical  (\bh{} masses and spins) and extrinsic  (position of the source in the sky of the
detector and orientation of the binary) parameters. Since we are only concerned with line-of-sight orientation effects, we will  assume that the detector is \emph{optimally oriented} and $\vec\xi_s = \vec\xi_u$, and in particular $(\theta_s,\phi_s) = (\theta_u,\phi_u)$. As a consequence, once a source is selected, $\epsilon$ will only depend on the location of the source in the sky, i.e.  $\epsilon=\epsilon(\theta,\phi)$ as implied by Eq.~(\ref{eq:recievedSignal2}). 
In section~\ref{subsec:bank_maximization}, we will investigate, in addition to orientation effects, the consequences of allowing differences in
intrinsic physical parameters between the signal and the templates.

\begin{figure}[tb]
\centering
\vbox{
    \includegraphics[width=.70\linewidth]{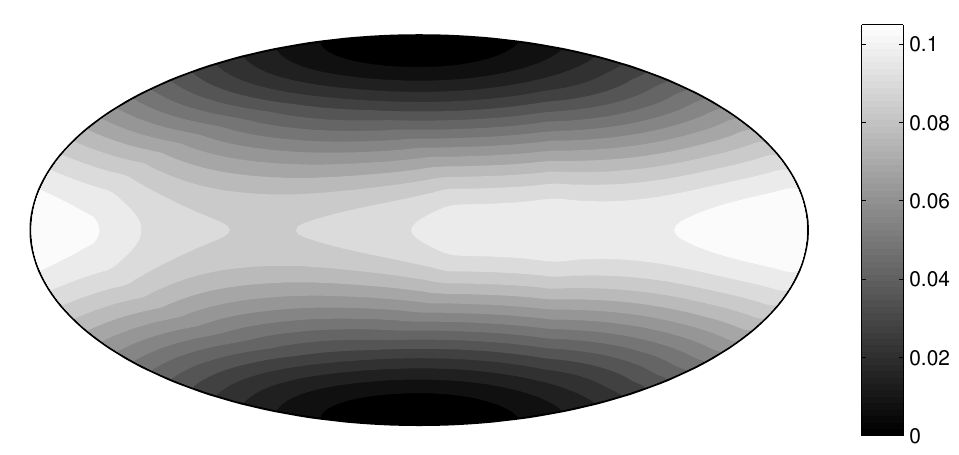}
    \includegraphics[width=.70\linewidth]{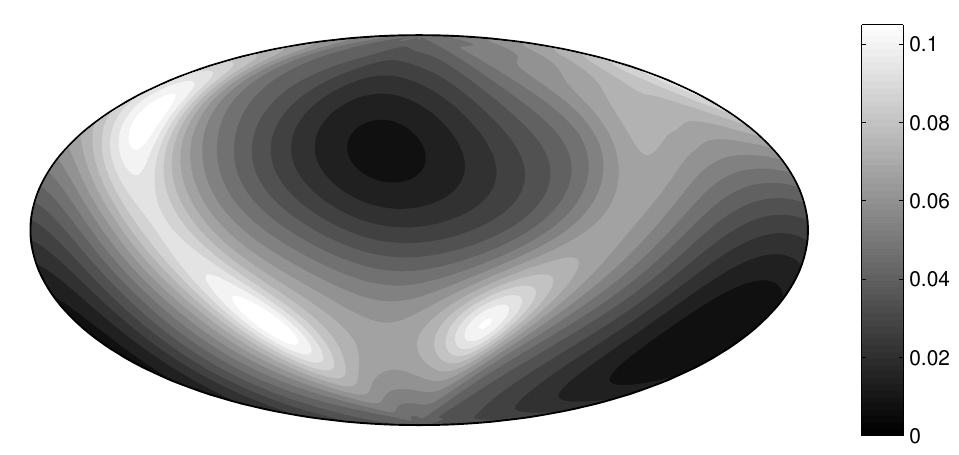}
}
\caption{Representative sky-maps of mismatches for a \bbh{} with mass ratio $q=m_1/m_2=4$. The top panel shows a non-precessing binary (case Q06) in which the template contains only the (2,2) mode.  In the bottom panel, the precessing case P19 is shown using a template that contains all the $m$-modes for $\ell=2$.}
\label{fig:alignedMaps}
\end{figure}

We will mostly present results as sky-maps, and investigate how $\epsilon(\theta,\phi)$  depends on the particular content of higher modes in the template.
We construct  sky-maps with $61\times61$ pixels; each pixel ``colored'' with its value of $\epsilon$. Figure~\ref{fig:alignedMaps} gives an example of two sky-maps of mismatches for a \bbh{} with mass ratio $q=4$. The top panel shows a non-precessing binary (case Q06 in Table~\ref{tab:NR}) in which the template contains only the (2,2) mode.  The bottom panel shows a precessing binary (case P19 in Table~\ref{tab:NR}).
The template in this case contains all the $m$-modes for $\ell=2$.  
 Notice that, the mismatch is smaller around the poles in the non-precessing case. On the other hand, for the precessing case,
the region where the mismatches are smaller have been shifted due to the precession of the system.

In addition to the sky-maps, given a mismatch tolerance $\epsilon_{\rm mx}$, we label a pixel \emph{on} if $\epsilon < \epsilon_{\rm mx}$ and \emph{off} otherwise. With the on-pixels, we estimate the covering factor $\tau(\epsilon_{\rm mx})$ of a template as the fraction of the sky where $\epsilon < \epsilon_{\rm mx}$. This means that a template that cannot detect any signals would have a
covering factor of 0, and a template that matches all signals would
have a covering factor of 1.  From now on, when stating covering
factors, we will just use the values of $\tau$, understanding that $\epsilon_{\rm
mx}$ is implied. Unless explicitly specified, we use $\epsilon_{mx} =
0.03$ in the remainder of the paper, a value commonly used by the \gw{} data analysis
community. For instance, with $\epsilon_{mx} =
0.03$, the covering factors for the cases in Figure~\ref{fig:alignedMaps} are $\tau = 0.177$ (top) and $\tau = 0.242$ (bottom). 
Furthermore, a value  $\epsilon_{mx} = 0.03$ corresponds to a 3\% loss
of \snr{} or equivalently a 10\% loss of detection volume as implied by Eq.~(\ref{eq:volume}).

It should not be surprising that the mode content in a template has an impact on its mismatch with the signal, and thus its covering factor. Our goal is then to find the optimal template $u_N$ that yields a target covering factor $\tau_{\rm tg}$. We use a \emph{greedy} algorithm for this purpose. The end result is a  sequence of templates  $\lbrace u_n\rbrace$ built from a set of modes $\lbrace w_n\rbrace$ as in Eq.~(\ref{eq:templates}), with $n = 1\dots N$. At each step in the sequence, a template $u_n$ is locally optimal in its ability for improving sky coverage and reaching $\tau_{\rm tg}$. Furthermore, the template sequence is ordered in a \emph{hierarchy} with respect to their sky coverage capabilities. The procedure we use to obtain this hierarchy involves three stages. 

\emph{First Stage:} The first template $u_1$ is selected to be the dominant (2,2) mode. The covering factor for this first template is denoted by $\tau_1$. The second template $u_2$ is obtained from a superposition of the previous template, $u_1$, and  a mode $w_i$ selected from the remaining 31 modes. This yields 31 potential candidates for $u_2$.  We promote to template $u_2$ the candidate that delivers the best improvement in sky coverage over the value $\tau_1$. The third template in the hierarchy is constructed in a similar fashion. That is, $u_3 = u_2$ plus one of the 30 remaining $w_i$ modes. Among the 30 candidates for $u_3$, we keep the one with the largest improvement in sky coverage over the value $\tau_2$. This process is repeated until we obtain a template, $u_N$, such that $\tau_N \ge \tau_{\rm tg}$. 

\emph{Second Stage:} Because the procedure to get $\lbrace u_n\rbrace$  is locally optimal, it is not guaranteed that the sequence has the minimum possible number of steps to reach  $\tau_{\rm tg}$, in other words, that the template $u_N$ has the minimum  number of modes. Next is to identify all modes in $u_N$ that could be excluded and  still have $\tau_N \ge \tau_{\rm tg}$. 
Starting with $u_N$,  
we construct templates 
\begin{equation}
 u^*_n = u_N - w_n = \sum_{i=1}^N w_i - w_n =  \sum_{i\ne n}^N w_i
\end{equation}
for $n=1\dots N$.
We then select the  $u^*_n$ for which $\tau^*_n \ge \tau_{\rm tg}$. If there is more than one template $u^*_n$, we single out the case in which the highest mode was subtracted. That mode is then remove from the template $u_N$ and set $\bar u_{N-1} = u^*_n$, which contains $N-1$ modes.
The same process is applied now to $\bar u_{N-1}$. That is, we construct templates
\begin{equation}
 u^*_n = \bar u_{N-1} - w_n = \sum_{i=1}^{N-1} w_i - w_n =  \sum_{i\ne n}^{N-1} w_i
\end{equation}
for $n=1\dots N-1$. The outcome is a new template $\bar u_{N-2}$. The process is repeated $L$ times until we are not able to construct templates $u^*_n$ such that $\tau^*_n \ge \tau_{\rm tg}$.  The resulting $\bar u_M$, with $M=N-L$, is the template with the minimum number of modes having a covering factor $\tau_{\rm tg}$.

\emph{Third Stage:} With the template $\bar u_M$ at hand, we apply a greedy algorithm to re-construct the hierarchy but now in reverse order; that is, 
starting with $\bar u_M$ we obtain $\bar u_{M-1}$ as the template that decreases $\bar \tau_{M}$ 
 the least. We continue until we reach a template made out of a single mode.

Tables~\ref{tab:greedy}, ~\ref{tab:greedy2}  and~\ref{tab:greedy3}  show how our greedy algorithm builds a hierarchy for the P17, P19 and P20 cases, respectively. 
The top row labels the $(\ell,m)$ modes, and the left column denotes the covering factor $\tau$. The target covering factor is $\tau_{tg} = 1$.  
For a given $\tau$, the Xs in the row indicated the modes that were included in the template.
The Os denote the mode from the previous step that was removed.
The first horizontal line denotes the end of the \emph{first stage} in the greedy algorithm. The rows between the first and second line are the steps involved in the \emph{second stage}. Finally, below the second line are the templates 
in hierarchy one is seeking, in decreasing order of sky coverage.

\begin{table*}
 \begin{tabular}{l|ccccccccccccccccccccc}
\multicolumn{1}{l|}{\multirow{2}{*}{\backslashbox{$\tau$}{$l,m$}}}       &2 &2 &3 &3 &2 &3 &3 &2 &2 &3 &4 &4 &4 &4 &4 &5 &4 &5 &5 &3 &3\\
\multicolumn{1}{l|}{}       &2 &0 &1 &3 &-2&-1&0 &1 &-1&-3&-3&3 &-2&0 &-4&3 &2 &-4&-1&-2&2\\
\hline
\hline
0.019 &X &  &  &  &  &  &  &  &  &  &  &  &  &  &  &  &  &  &  &  & \\
0.128 &X &X &  &  &  &  &  &  &  &  &  &  &  &  &  &  &  &  &  &  & \\
0.213 &X &X &X &  &  &  &  &  &  &  &  &  &  &  &  &  &  &  &  &  & \\
0.259 &X &X &X &X &  &  &  &  &  &  &  &  &  &  &  &  &  &  &  &  & \\
0.298 &X &X &X &X &X &  &  &  &  &  &  &  &  &  &  &  &  &  &  &  & \\
0.361 &X &X &X &X &X &X &  &  &  &  &  &  &  &  &  &  &  &  &  &  & \\
0.404 &X &X &X &X &X &X &X &  &  &  &  &  &  &  &  &  &  &  &  &  & \\
0.473 &X &X &X &X &X &X &X &X &  &  &  &  &  &  &  &  &  &  &  &  & \\
0.787 &X &X &X &X &X &X &X &X &X &  &  &  &  &  &  &  &  &  &  &  & \\
0.904 &X &X &X &X &X &X &X &X &X &X &  &  &  &  &  &  &  &  &  &  & \\
0.912 &X &X &X &X &X &X &X &X &X &X &X &  &  &  &  &  &  &  &  &  & \\
0.919 &X &X &X &X &X &X &X &X &X &X &X &X &  &  &  &  &  &  &  &  & \\
0.922 &X &X &X &X &X &X &X &X &X &X &X &X &X &  &  &  &  &  &  &  & \\
0.925 &X &X &X &X &X &X &X &X &X &X &X &X &X &X &  &  &  &  &  &  & \\
0.926 &X &X &X &X &X &X &X &X &X &X &X &X &X &X &X &  &  &  &  &  & \\
0.927 &X &X &X &X &X &X &X &X &X &X &X &X &X &X &X &X &  &  &  &  & \\
0.928 &X &X &X &X &X &X &X &X &X &X &X &X &X &X &X &X &X &  &  &  & \\
0.929 &X &X &X &X &X &X &X &X &X &X &X &X &X &X &X &X &X &X &  &  & \\
0.930 &X &X &X &X &X &X &X &X &X &X &X &X &X &X &X &X &X &X &X &  & \\
0.931 &X &X &X &X &X &X &X &X &X &X &X &X &X &X &X &X &X &X &X &X & \\
1.000 &X &X &X &X &X &X &X &X &X &X &X &X &X &X &X &X &X &X &X &X &X\\
\hline
\multicolumn{1}{l|}{\multirow{2}{*}{\backslashbox{$\tau$}{$l,m$}}}       &2 &2 &3 &3 &2 &3 &3 &2 &2 &3 &4 &4 &4 &4 &4 &5 &4 &5 &5 &3 &3\\
\multicolumn{1}{l|}{}       &2 &0 &1 &3 &-2&-1&0 &1 &-1&-3&-3&3 &-2&0 &-4&3 &2 &-4&-1&-2&2\\
\hline
1.000 &X &X &X &X &X &X &X &X &X &X &X &X &X &X &X &O &X &X &X &X &X\\
1.000 &X &X &X &X &X &X &X &X &X &X &X &X &X &X &X &  &X &X &O &X &X\\
1.000 &X &X &X &X &X &X &X &X &X &X &X &X &X &X &X &  &X &O &  &X &X\\
1.000 &X &X &X &X &X &X &X &X &X &X &X &O &X &X &X &  &X &  &  &X &X\\
1.000 &X &X &X &X &X &X &X &X &X &X &X &  &X &X &X &  &O &  &  &X &X\\
1.000 &X &X &X &X &X &X &X &X &X &X &X &  &X &O &X &  &  &  &  &X &X\\
1.000 &X &X &X &X &X &X &X &X &X &X &X &  &O &  &X &  &  &  &  &X &X\\
1.000 &X &X &X &X &X &X &X &X &X &X &O &  &  &  &X &  &  &  &  &X &X\\
\hline
\multicolumn{1}{l|}{\multirow{2}{*}{\backslashbox{$\tau$}{$l,m$}}}       &2 &2 &3 &3 &2 &3 &3 &2 &2 &3 &4 &4 &4 &4 &4 &5 &4 &5 &5 &3 &3\\
\multicolumn{1}{l|}{}       &2 &0 &1 &3 &-2&-1&0 &1 &-1&-3&-3&3 &-2&0 &-4&3 &2 &-4&-1&-2&2\\
\hline
1.000 &X &X &X &X &X &X &X &X &X &X &  &  &  &  &O &  &  &  &  &X &X\\
0.887 &X &X &O &X &X &X &X &X &X &X &  &  &  &  &  &  &  &  &  &X &X\\
0.825 &X &X &  &X &X &O &X &X &X &X &  &  &  &  &  &  &  &  &  &X &X\\
0.705 &X &X &  &X &X &  &X &X &X &O &  &  &  &  &  &  &  &  &  &X &X\\
0.592 &X &X &  &X &X &  &X &X &X &  &  &  &  &  &  &  &  &  &  &X &O\\
0.619 &X &X &  &X &X &  &X &X &X &  &  &  &  &  &  &  &  &  &  &O & \\
0.520 &X &X &  &X &O &  &X &X &X &  &  &  &  &  &  &  &  &  &  &  & \\
0.368 &X &X &  &X &  &  &O &X &X &  &  &  &  &  &  &  &  &  &  &  & \\
0.238 &X &X &  &X &  &  &  &X &O &  &  &  &  &  &  &  &  &  &  &  & \\
0.174 &X &X &  &X &  &  &  &O &  &  &  &  &  &  &  &  &  &  &  &  & \\
0.128 &X &X &  &O &  &  &  &  &  &  &  &  &  &  &  &  &  &  &  &  & \\
0.031 &O &X &  &  &  &  &  &  &  &  &  &  &  &  &  &  &  &  &  &  & 
\end{tabular}

\caption{\textbf{P17 Hierarchy}:
    Table showing how our method for constructing the mode hierarchy works for the case P17.  The top row labels the $(\ell,m)$ modes,
   and the left column denotes the covering factor $\tau$. The target covering factor is $\tau_{tg} = 1$.  For a given $\tau$, the Xs in the row indicated the modes that were included in the template. The Os denote the mode from the previous step that was removed. The first horizontal line denotes the end of the \emph{first stage} in the greedy algorithm. The rows between the first and second line are the steps involved in the \emph{second stage}. Finally, below the second line are the templates 
in hierarchy one is seeking, in decreasing order of sky coverage.}
\label{tab:greedy}
\end{table*}

\begin{table*}
 \begin{tabular}{l|cccccccccccccccccccc}
\multicolumn{1}{l|}{\multirow{2}{*}{\backslashbox{$\tau$}{$l,m$}}}       &2 &2 &3 &2 &3 &3 &3 &2 &2 &3 &4 &4 &4 &4 &4 &5 &4 &4 &3 &3\\
\multicolumn{1}{l|}{}       &2 &0 &1 &-2&-3&-1&0 &-1& 1& 3& 3&-3& 2&0 &-2&3 &-4& 4& 2&-2\\
\hline
\hline
0.001 &X &  &  &  &  &  &  &  &  &  &  &  &  &  &  &  &  &  &  &  \\
0.081 &X &X &  &  &  &  &  &  &  &  &  &  &  &  &  &  &  &  &  &  \\
0.167 &X &X &X &  &  &  &  &  &  &  &  &  &  &  &  &  &  &  &  &  \\
0.212 &X &X &X &X &  &  &  &  &  &  &  &  &  &  &  &  &  &  &  &  \\
0.280 &X &X &X &X &X &  &  &  &  &  &  &  &  &  &  &  &  &  &  &  \\
0.340 &X &X &X &X &X &X &  &  &  &  &  &  &  &  &  &  &  &  &  &  \\
0.365 &X &X &X &X &X &X &X &  &  &  &  &  &  &  &  &  &  &  &  &  \\
0.410 &X &X &X &X &X &X &X &X &  &  &  &  &  &  &  &  &  &  &  &  \\
0.783 &X &X &X &X &X &X &X &X &X &  &  &  &  &  &  &  &  &  &  &  \\
0.915 &X &X &X &X &X &X &X &X &X &X &  &  &  &  &  &  &  &  &  &  \\
0.924 &X &X &X &X &X &X &X &X &X &X &X &  &  &  &  &  &  &  &  &  \\
0.931 &X &X &X &X &X &X &X &X &X &X &X &X &  &  &  &  &  &  &  &  \\
0.933 &X &X &X &X &X &X &X &X &X &X &X &X &X &  &  &  &  &  &  &  \\
0.935 &X &X &X &X &X &X &X &X &X &X &X &X &X &X &  &  &  &  &  &  \\
0.937 &X &X &X &X &X &X &X &X &X &X &X &X &X &X &X &  &  &  &  &  \\
0.938 &X &X &X &X &X &X &X &X &X &X &X &X &X &X &X &X &  &  &  &  \\
0.939 &X &X &X &X &X &X &X &X &X &X &X &X &X &X &X &X &X &  &  &  \\
0.940 &X &X &X &X &X &X &X &X &X &X &X &X &X &X &X &X &X &X &  &  \\
0.943 &X &X &X &X &X &X &X &X &X &X &X &X &X &X &X &X &X &X &X &  \\
1.000 &X &X &X &X &X &X &X &X &X &X &X &X &X &X &X &X &X &X &X &X \\
\hline
\multicolumn{1}{l|}{\multirow{2}{*}{\backslashbox{$\tau$}{$l,m$}}}       &2 &2 &3 &2 &3 &3 &3 &2 &2 &3 &4 &4 &4 &4 &4 &5 &4 &4 &3 &3\\
\multicolumn{1}{l|}{}       &2 &0 &1 &-2&-3&-1&0 &-1& 1& 3& 3&-3& 2&0 &-2&3 &-4& 4& 2&-2\\
\hline
1.000 &X &X &X &X &X &X &X &X &X &X &X &X &X &X &X &O &X &X &X &X \\
1.000 &X &X &X &X &X &X &X &X &X &X &X &X &X &X &X &  &X &O &X &X \\
1.000 &X &X &X &X &X &X &X &X &X &X &O &X &X &X &X &  &X &  &X &X \\
1.000 &X &X &X &X &X &X &X &X &X &X &  &X &O &X &X &  &X &  &X &X \\
1.000 &X &X &X &X &X &X &X &X &X &X &  &X &  &O &X &  &X &  &X &X \\
1.000 &X &X &X &X &X &X &X &X &X &X &  &X &  &  &O &  &X &  &X &X \\
1.000 &X &X &X &X &X &X &X &X &X &X &  &O &  &  &  &  &X &  &X &X \\
\hline
\multicolumn{1}{l|}{\multirow{2}{*}{\backslashbox{$\tau$}{$l,m$}}}       &2 &2 &3 &2 &3 &3 &3 &2 &2 &3 &4 &4 &4 &4 &4 &5 &4 &4 &3 &3\\
\multicolumn{1}{l|}{}       &2 &0 &1 &-2&-3&-1&0 &-1& 1& 3& 3&-3& 2&0 &-2&3 &-4& 4& 2&-2\\
\hline
1.000 &X &X &X &X &X &X &X &X &X &X &  &  &  &  &  &  &O &  &X &X \\
0.907 &X &X &X &X &X &X &X &X &X &X &  &  &  &  &  &  &  &  &O &X \\
0.915 &X &X &X &X &X &X &X &X &X &X &  &  &  &  &  &  &  &  &  &O \\
0.806 &X &X &O &X &X &X &X &X &X &X &  &  &  &  &  &  &  &  &  &  \\
0.763 &X &X &  &X &X &O &X &X &X &X &  &  &  &  &  &  &  &  &  &  \\
0.605 &X &X &  &X &O &  &X &X &X &X &  &  &  &  &  &  &  &  &  &  \\
0.466 &X &X &  &O &  &  &X &X &X &X &  &  &  &  &  &  &  &  &  &  \\
0.323 &X &X &  &  &  &  &O &X &X &X &  &  &  &  &  &  &  &  &  &  \\
0.202 &X &X &  &  &  &  &  &O &X &X &  &  &  &  &  &  &  &  &  &  \\
0.125 &X &X &  &  &  &  &  &  &O &X &  &  &  &  &  &  &  &  &  &  \\
0.081 &X &X &  &  &  &  &  &  &  &O &  &  &  &  &  &  &  &  &  &  \\
0.023 &O &X &  &  &  &  &  &  &  &  &  &  &  &  &  &  &  &  &  &  
\end{tabular}

\caption{\textbf{P19 Hierarchy}:
    Same as in Table~\ref{tab:greedy} but for the P19 case.}
\label{tab:greedy2}
\end{table*}

\begin{table*}
 \begin{tabular}{l|ccccccccccccccc}
\multicolumn{1}{l|}{\multirow{2}{*}{\backslashbox{$\tau$}{$l,m$}}}       &2 &3 &4 &5 &5 &5 &5 &5 &5 &2 &3 &2 &3 &2 &2 \\
\multicolumn{1}{l|}{}       &2 &3 &4 &5 & 4& 3& 2& 1& 0& 0& 1&-2&-3&-1& 1\\
\hline
\hline
0.155 &X &  &  &  &  &  &  &  &  &  &  &  &  &  &  \\
0.506 &X &X &  &  &  &  &  &  &  &  &  &  &  &  &  \\
0.588 &X &X &X &  &  &  &  &  &  &  &  &  &  &  &  \\
0.597 &X &X &X &X &  &  &  &  &  &  &  &  &  &  &  \\
0.591 &X &X &X &X &X &  &  &  &  &  &  &  &  &  &  \\
0.573 &X &X &X &X &X &X &  &  &  &  &  &  &  &  &  \\
0.548 &X &X &X &X &X &X &X &  &  &  &  &  &  &  &  \\
0.542 &X &X &X &X &X &X &X &X &  &  &  &  &  &  &  \\
0.535 &X &X &X &X &X &X &X &X &X &  &  &  &  &  &  \\
0.534 &X &X &X &X &X &X &X &X &X &X &  &  &  &  &  \\
0.555 &X &X &X &X &X &X &X &X &X &X &X &  &  &  &  \\
0.563 &X &X &X &X &X &X &X &X &X &X &X &X &  &  &  \\
0.747 &X &X &X &X &X &X &X &X &X &X &X &X &X &  &  \\
0.950 &X &X &X &X &X &X &X &X &X &X &X &X &X &X &  \\
1.000 &X &X &X &X &X &X &X &X &X &X &X &X &X &X &X \\
\hline
\multicolumn{1}{l|}{\multirow{2}{*}{\backslashbox{$\tau$}{$l,m$}}}       &2 &3 &4 &5 &5 &5 &5 &5 &5 &2 &3 &2 &3 &2 &2 \\
\multicolumn{1}{l|}{}       &2 &3 &4 &5 & 4& 3& 2& 1& 0& 0& 1&-2&-3&-1& 1\\
\hline
1.000 &X &X &X &O &X &X &X &X &X &X &X &X &X &X &X \\
1.000 &X &X &X &  &O &X &X &X &X &X &X &X &X &X &X \\
1.000 &X &X &X &  &  &O &X &X &X &X &X &X &X &X &X \\
1.000 &X &X &X &  &  &  &O &X &X &X &X &X &X &X &X \\
1.000 &X &X &X &  &  &  &  &O &X &X &X &X &X &X &X \\
1.000 &X &X &X &  &  &  &  &  &O &X &X &X &X &X &X \\
1.000 &X &X &O &  &  &  &  &  &  &X &X &X &X &X &X \\
\hline
\multicolumn{1}{l|}{\multirow{2}{*}{\backslashbox{$\tau$}{$l,m$}}}       &2 &3 &4 &5 &5 &5 &5 &5 &5 &2 &3 &2 &3 &2 &2 \\
\multicolumn{1}{l|}{}       &2 &3 &4 &5 & 4& 3& 2& 1& 0& 0& 1&-2&-3&-1& 1\\
\hline
1.000 &X &X &  &  &  &  &  &  &  &X &O &X &X &X &X \\
0.957 &X &X &  &  &  &  &  &  &  &O &  &X &X &X &X \\
0.748 &X &X &  &  &  &  &  &  &  &  &  &X &X &X &O \\
0.565 &X &X &  &  &  &  &  &  &  &  &  &X &X &O &  \\
0.406 &X &X &  &  &  &  &  &  &  &  &  &X &O &  &  \\
0.506 &X &X &  &  &  &  &  &  &  &  &  &O &  &  &  \\
0.155 &X &O &  &  &  &  &  &  &  &  &  &  &  &  &  \\
\end{tabular}

\caption{\textbf{P20 Hierarchy}:
    Same as in Table~\ref{tab:greedy} but for the P20 case.}
\label{tab:greedy3}
\end{table*}

\section{Results}
\label{sec:results}

A sense of the effectiveness of the mode (2,2) in covering the sky is given in the second to last column of Table~\ref{tab:NR} where we report the covering factor of just the (2,2) mode, given by $\tau_{2,2}$. 
As long as $q \le 1.5$, the (2,2) mode will effectively cover the entire sky for non-spinning and non-precessing \bh{s} (see Q and S series). 
Even in the case of moderate precession (e.g. cases P01and P02) the quadruple mode is sufficient. At the same time, it is evident from the Q-series that as $q$ increases
the effectiveness of a (2,2)-only template quickly deteriorates. For $q= 4$, the covering factor is already $\tau = 0.177$, reaching 0.097 for $q=10$.
In the last column of Table~\ref{tab:NR}, we report the covering factor $\tau_{2,*}$ when the template includes all $|m| \le \ell = 2$ modes. Notice that for the non-precessing binaries 
(series Q and S), the inclusion of these modes does not translate into a significant improvement of sky coverage. 
For the precessing systems, $\tau_{2,2}$ is not a consistent description of the coverage, because it depends 
on the fixed extraction frame.  A frame independent choice is $\tau_{2,*}$.  Typically, $\tau_{2,*}$ improves coverage by at least a factor of 2 for the precessing cases, 
and in the highly precessing cases, by a significant amount more.

\subsection{Q-series}

\begin{figure}[tb]
\centering
\vbox{
    \includegraphics[width=.70\linewidth,angle=270]{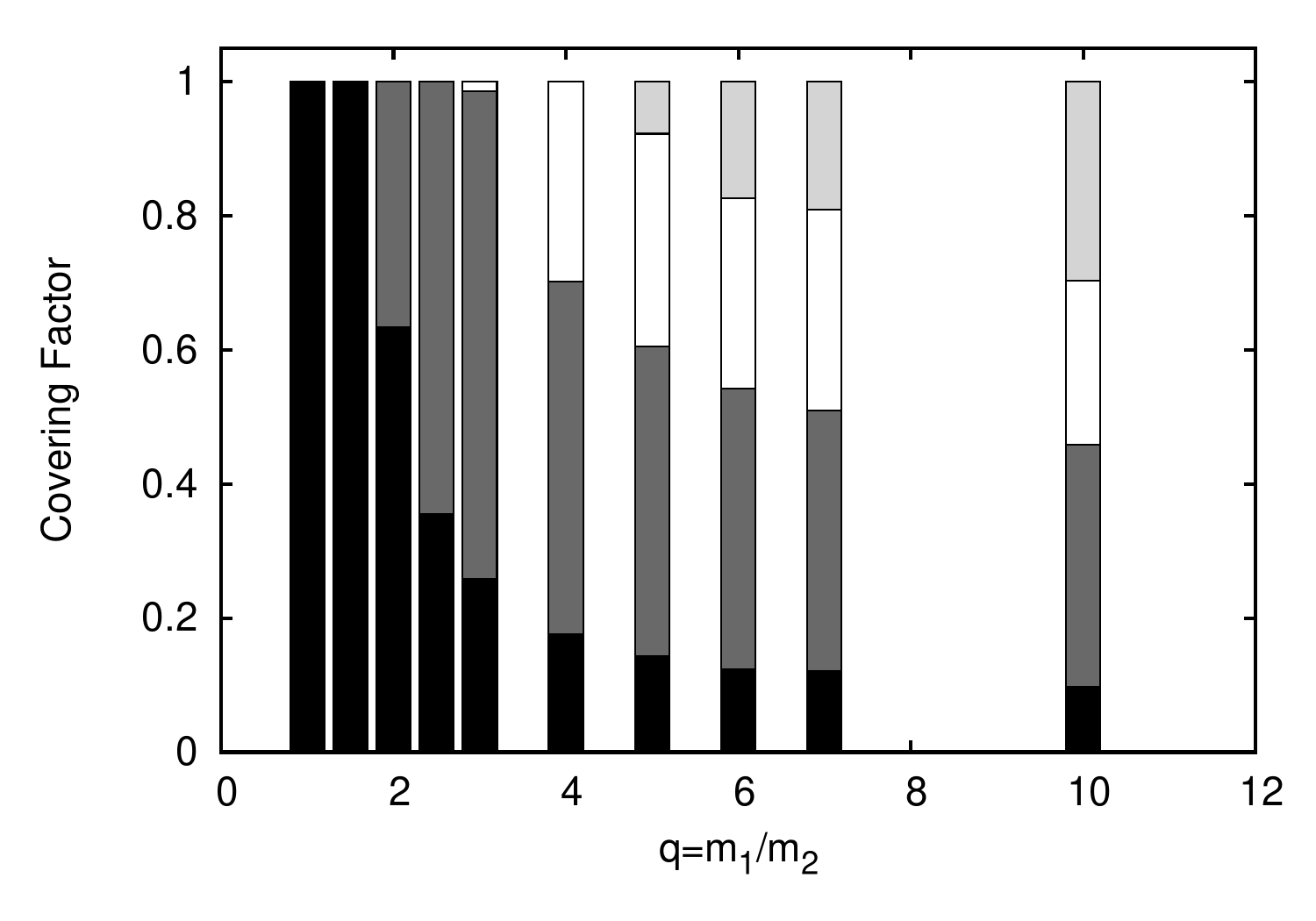}
    \includegraphics[width=.70\linewidth,angle=270]{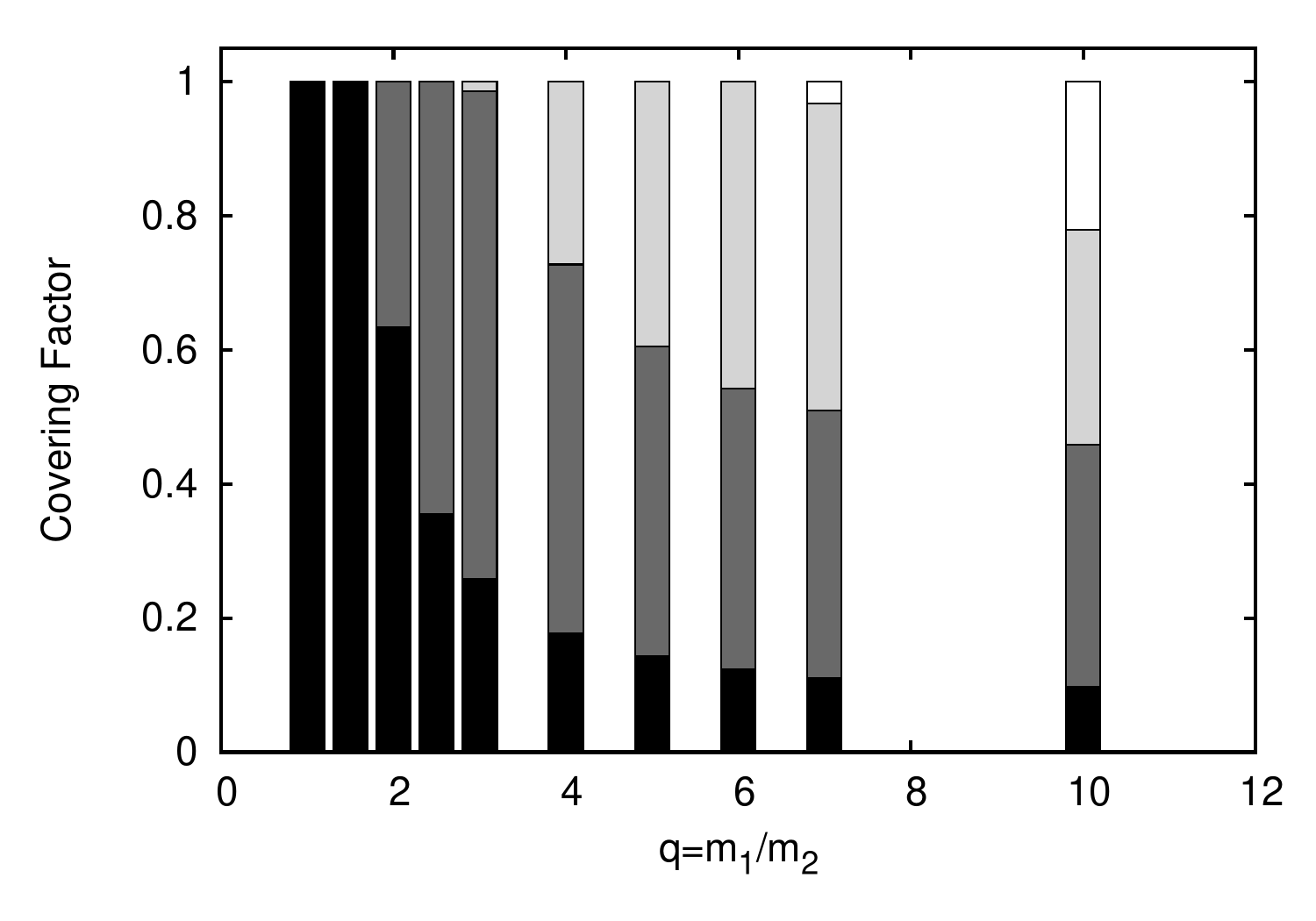}
}
\caption{Covering factors as a function of mass ratio for the Q-series. Each portion of the bar chart
represents the contribution to the covering factor by adding a mode. In each bar, bottom to top is the order used 
in adding modes. The gray scale is as follows: black  (2,2) mode, gray (3,3) mode, white (4,4) mode and light gray (2,1) mode. 
Top panel shows the case in which modes are added according to their energy emission strength. The bottom panel shows
the hierarchy order obtained by our greedy algorithm, which translates into reversing the order of the (4,4) and (2,1) modes.
}
\label{fig:massratiobar}
\end{figure}

We first analyze the impact of adding other modes to the quadrupolar templates for the Q-series. 
To anticipate what to expect from the greedy algorithm described in Sec.~\ref{sec:greedy}, 
we will consider first adding modes based only on their energy radiated; these are (3,3), (4,4) and (2,1) for the Q-series. 
Adding the (3,3) mode translates into a significant improvement as depicted in the top panel  in Figure~\ref{fig:massratiobar} 
where we show a bar chart of the covering factor as a function of the mass ratio. For each bar, black denotes the covering factor that is achieved
with only the (2,2) mode. The gray bar above black is the enhancement that one gets by considering a template with (2,2) + (3,3). One sees that now 
full coverage is obtained for  $q \le 3$ and substantial coverage improvement of $0.5 \le \tau \le 0.75$ for $7 \ge q \ge 5$, and $\tau = 0.45$ for $q = 10$. 
Adding the next most energetic mode, the (4,4) mode, produces coverage above 0.8 for the remaining cases except $q=10$, 
as seen by the white bars in Figure~\ref{fig:massratiobar} top panel.
Full coverage for all the mass ratios considered is achieved when the (2,1) is added (light gray bars).

\begin{figure}[tb]
\centering
\vbox{
    \includegraphics[width=.70\linewidth]{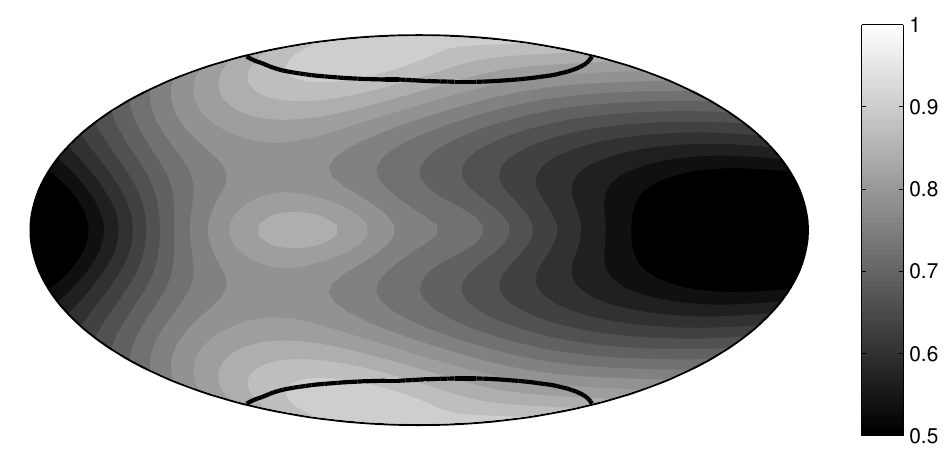}
    \includegraphics[width=.70\linewidth]{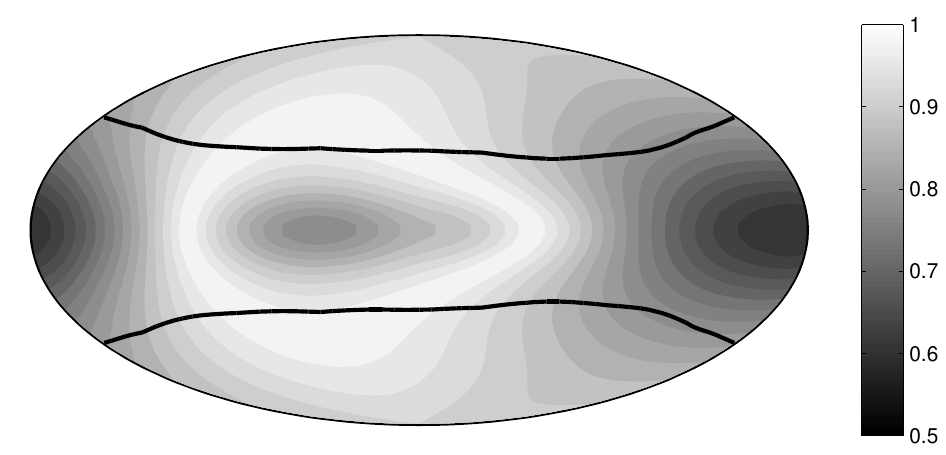}
    \includegraphics[width=.70\linewidth]{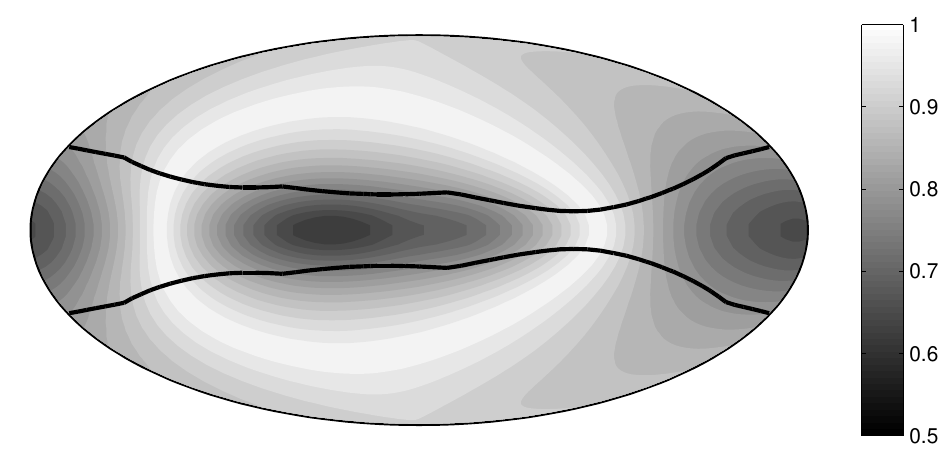}
    \includegraphics[width=.70\linewidth]{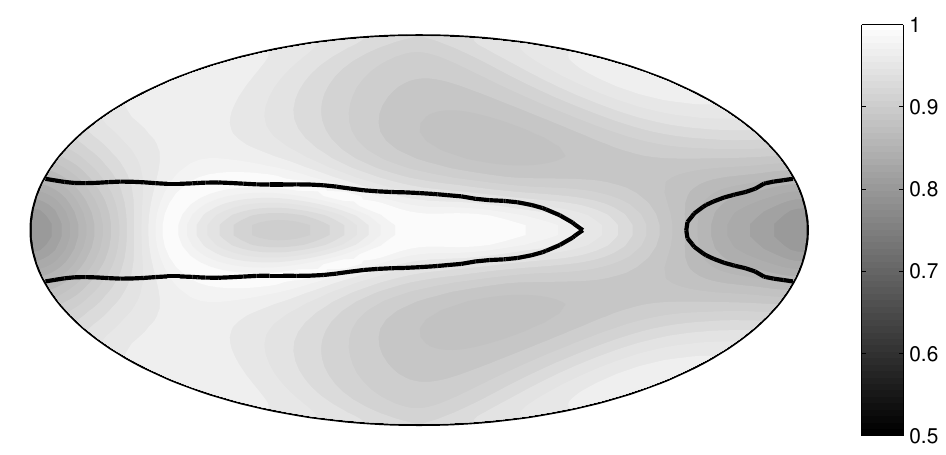}
}
\caption{
Percentage of total energy emitted in a given direction that is channeled through the modes present in a template for the $q=10$ case in the Q-series. 
From top to bottom,  the panels show the cases for templates (2,2), (2,2)+(3,3), (2,2)+(3,3)+(4,4), and (2,2)+(3,3)+(2,1), respectively. 
The black lines in each panel denote mismatches $\epsilon =0.03$ between the template and the signal.
Regions containing the poles have mismatches values $\epsilon < 0.03$.  
}
\label{fig:q10Eradmaps}
\end{figure}

The ordering based on energy used for the Q-series gets slightly modified if one uses our greedy algorithm. The method reveals that
adding the (2,1) mode before the (4,4) mode is more optimal in reaching full coverage, as seen in the bottom panel of Figure~\ref{fig:massratiobar}.  
The reason why the greedy algorithm favors the (2,1)
mode for $q > 4$ is because, although the mode is less energetic, it channels energy into regions of 
the sky not covered by the (2,2)+(3,3) template more effectively than the (4,4) mode.  
This can be seen in Figure~\ref{fig:q10Eradmaps} for the $q=10$ case in the Q-series.
Each of the panels in this figure shows the percentage of total energy emitted in 
a given direction that is channeled through the modes present in a template. 
From top to bottom,  the panels show the cases for templates (2,2), (2,2)+(3,3), (2,2)+(3,3)+(4,4), and (2,2)+(3,3)+(2,1), respectively. 
The black lines
in each  denote mismatches $\epsilon =0.03$ between the template and the signal.
Regions containing the poles have mismatches values $\epsilon < 0.03$. 
Not surprisingly, most of the dark regions are contained in regions where $\epsilon > 0.03$ and light regions where $\epsilon < 0.03$;
that is, there is a correlation between the amount of energy
captured by the template and its  mismatch value. 
In other words, regions for which the template captures more than 80\% of the energy emitted in a given direction have $\epsilon \le 0.03$.
In these regions of low mismatch, the average percentage radiated in all four panels of Figure~\ref{fig:q10Eradmaps} is approximately 90\% overall.
It is then evident from the
bottom two panels in Figure~\ref{fig:q10Eradmaps}, that adding the (2,1) mode before the (4,4) yields a larger region where the 
template better captures the energy emitted.

Also interesting in Figure~\ref{fig:q10Eradmaps} is how the two regions with $\epsilon < 0.03$, one the north pole and the other at the south pole,
grow towards the equator as higher modes are added;  to the point that,
in the bottom panel, the two regions merge on the equator.
This can be explained by recalling that the grey shades in this figure depict the percent difference of the energy radiated in 
a signal containing all the modes and a template containing a subset of 
all the modes. Therefore, the bright white region along the equator in the bottom
panel indicates that the energy radiated in that region is well
captured by the template.  However, having an agreement between the signal and the template regarding the amount of energy radiated
is not the full story. A template and a signal that are energetically comparable do not necessarily have a low mismatch, as one can observe in the last panel  
in Figure~\ref{fig:q10Eradmaps} in the equator.

We have also found that there is a degree of degeneracy. For instance, in the $q=7$ case, 
adding to the template (2,2)+(3,3)+(2,1) the mode (4,4) or the mode (5,5)  accomplishes full coverage. That is, 
both modes contain enough power in comparable parts of the sky.

\subsection{S-series}

\begin{figure}[tb]
\centering
\vbox{
    \includegraphics[width=.70\linewidth,angle=270]{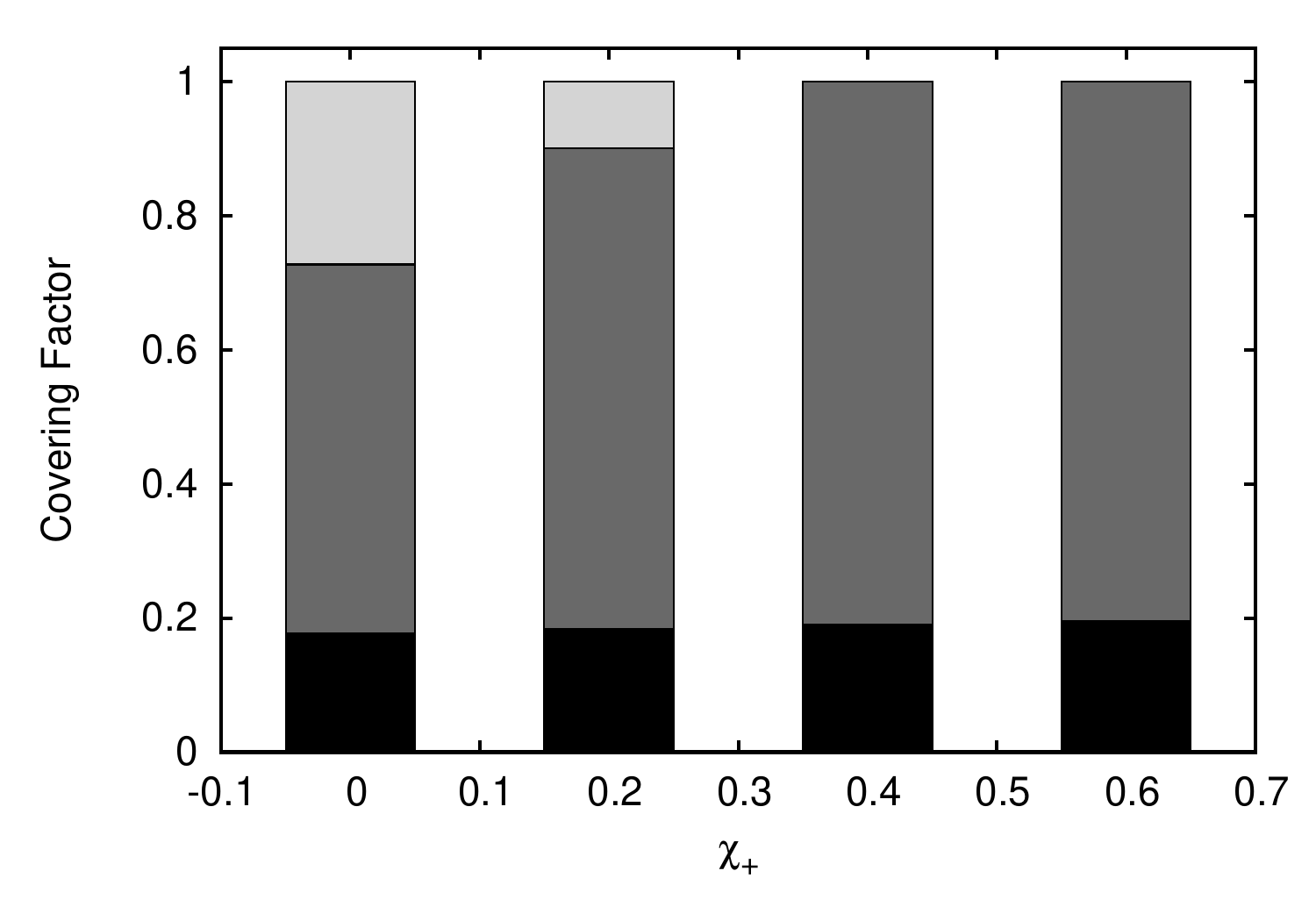}
    \includegraphics[width=.70\linewidth,angle=270]{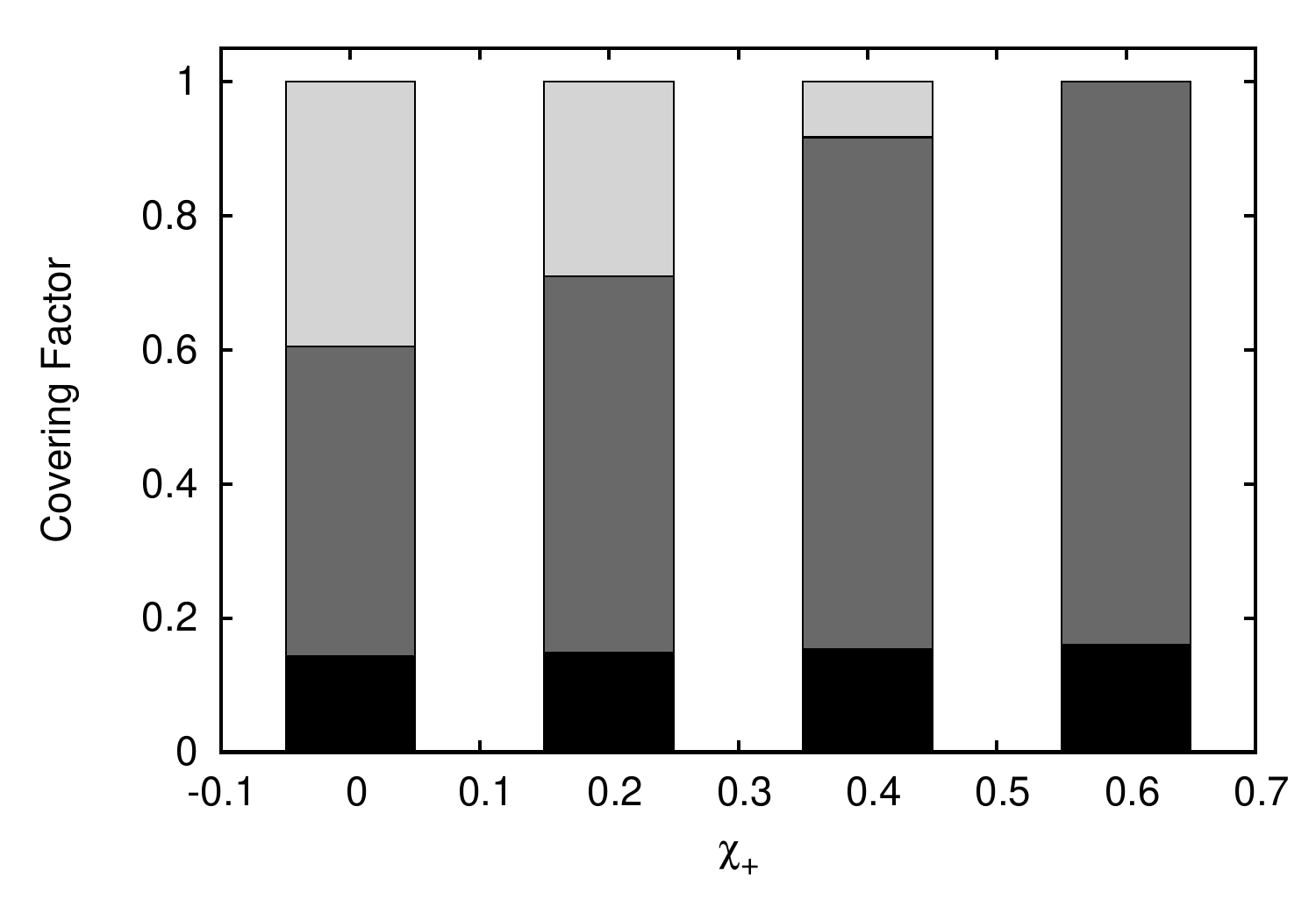}
    \includegraphics[width=.70\linewidth,angle=270]{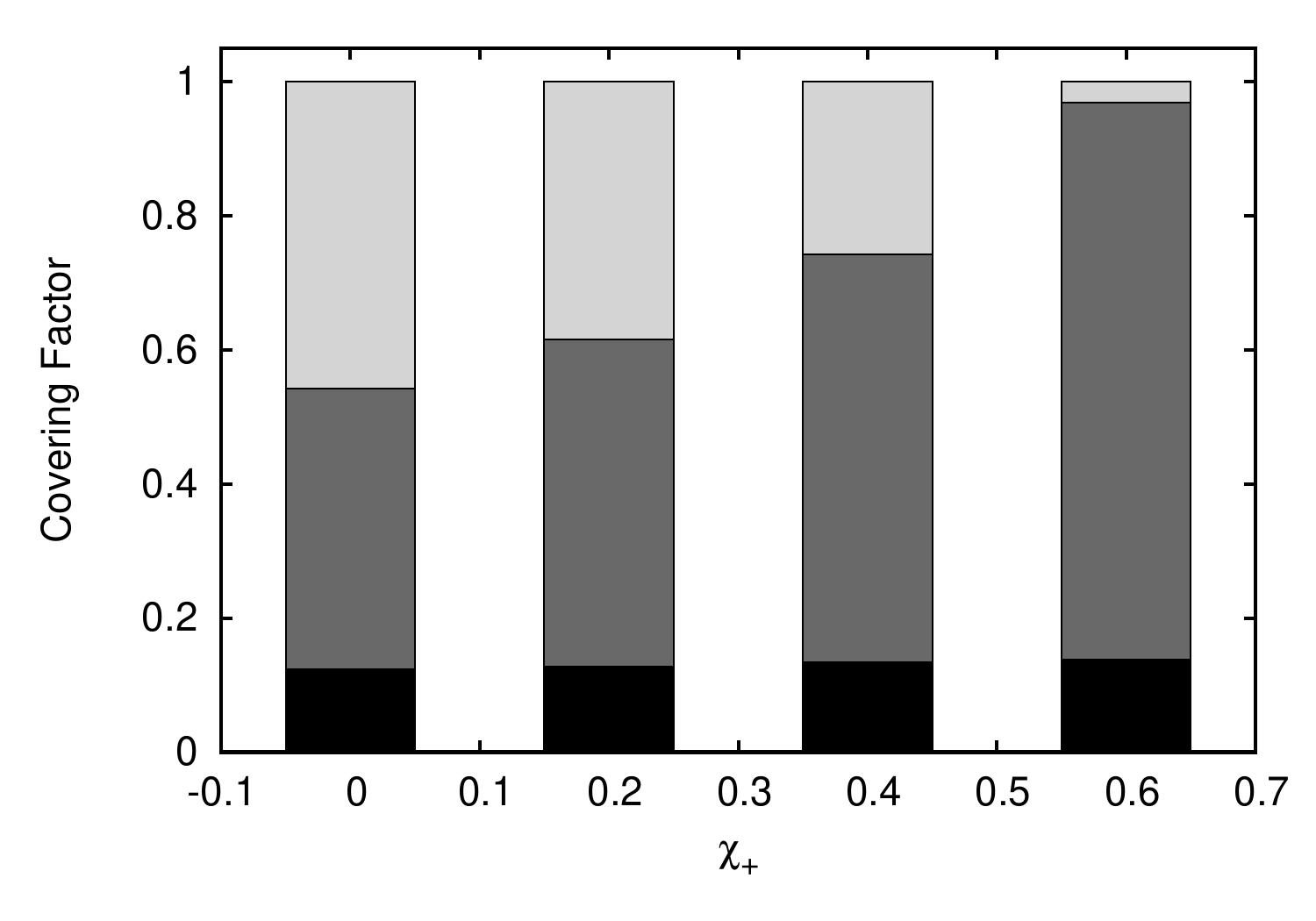}
}
\caption{  Covering factors versus spin for the S-series, $\chi_+ = a_+/m_+$, of the more massive \bh{.}  The less massive \bh{} has vanishing spin.  The top panel is $q=4$, center $q=5$, and bottom $q=6$.
The gray scale is as follows: black  (2,2) mode, gray (3,3) mode, and light gray (2,1) mode.
}
\label{fig:spinratiobar}
\end{figure}

Next to analyze is the S-series, consisting of \bbh{s} with spinning \bh{s} aligned with the orbital angular momentum, i.e. non-precessing binaries. 
As mentioned before, for low mass ratio $q \le 1.5$, the quadrupolar mode is able to get good coverage of the sky. 
We will focus then on the $q = 4$, 5 and 6 cases in Table~\ref{tab:NR}. These are binaries with only one of the \bh{s}  spinning. 
As with Figure~\ref{fig:massratiobar}, the bar charts in Figure~\ref{fig:spinratiobar} show the incremental effectiveness of adding modes to a template. In this case,
the covering factor is plotted as a function of the dimensionless spin parameter $\chi_+$.  The gray scale is: black (2,2) mode, gray (3,3) mode and 
light gray (2,1) mode. From top to bottom, the panels show the $q=4$, 5 and 6 cases, respectively.

The first thing to notice is that if only the (2,2) mode is used as a template, the covering factors (black bars) are
basically independent of the spin. In addition, the values of the covering factors for the (2,2) mode template decrease as the mass ratio increases. 
The big changes occur when one adds the (3,3) mode to the quadrupolar template. In some instances, one is able to increase the coverage to include the entire sky. 
It  is also interesting that the influence of this mode increases with the value of the spin, consistent with the energy of the (2,2) and (3,3) modes increasing with spin.
For example, in the $q=4$ case, a (2,2)+(3,3) template is able to saturate the sky for $\chi_+ \ge  0.4$. Finally, for those cases in which the (3,3) is not able to yield full sky coverage, i.e. low spin, high mass ratio binaries, adding the (2,1) mode completes the sky. 

\subsection{P-series}

The last series focused on precessing binaries. 
In this  series, our greedy algorithm used to construct hierarchies provided insights on the number of modes that will be needed for
detecting \gw{} from astrophysically realistic binaries. We applied the greedy algorithm to each of the 23 cases in the series. However, we show the steps that the 
algorithm takes for only the P17, P19 and P20 cases. 
The steps are shown in Tables~\ref{tab:greedy}, ~\ref{tab:greedy2}  and~\ref{tab:greedy3}, respectively.

A dramatic aspect found for these highly precessing binaries is that it takes a substantial number of modes 
to cover at least 60\% of the sky. For the P17 case, one sees from Table~\ref{tab:greedy} that one 
needs a template with 8 modes: $(2,\pm 2)$, $(2,\pm1)$, (2,0), (3,3), (3,-2) and (3,0). 
Similarly, for the P19 in Table~\ref{tab:greedy2}, the template contains the $(2,\pm 2)$, $(2,\pm1)$, (2,0), (3,3) and (3,0) modes. 
Finally, for the
P20 case,  as seen from Table~\ref{tab:greedy3}, the template involves the $(2,\pm 2)$ and $(3,\pm3)$ modes. 
The entire hierarchy for each case is listed below the second line in each of the Tables ordered in decreasing covering factor.
It is clear then that precession plays a big role on the outcome of the hierarchy, and in particular the
modes that are needed for full coverage of the sky (i.e. the row below the second  line in the tables). 

Finally, notice that the last mode kept by the greedy algorithm in the P17 and P19 cases (bottom of Tables~\ref{tab:greedy} and~\ref{tab:greedy2})
is the (2,0) mode, intend of the (2,2) mode one used to start building the hierarchy. On the other hand, 
for the P20 case (bottom of Table~\ref{tab:greedy3}) is the mode (2,2) that remains at the end.
The reason for this is difference is because P17 and P19 are both highly precessing binaries of long lasting duration.
In addition, we are using a fixed inertial frame for wave
extraction that is aligned with the initial angular momentum, the
precession moves the peak emission away from the fixed frame, and
the modes mix.  Therefore, even though we start with the (2,2) mode
for the algorithm, when we work backwards in Stage 3, the algorithm
finds that the (2,0) mode provides the best coverage for those two
cases.  However, this does not have a huge implication since both, the (2,0) and the (2,2) mode, provide poor coverage
as standalone modes,  $\le 3\%$ of the sky in these cases.  

\begin{figure}[tb]
\centering
\vbox{
    \includegraphics[width=.70\linewidth,angle=270]{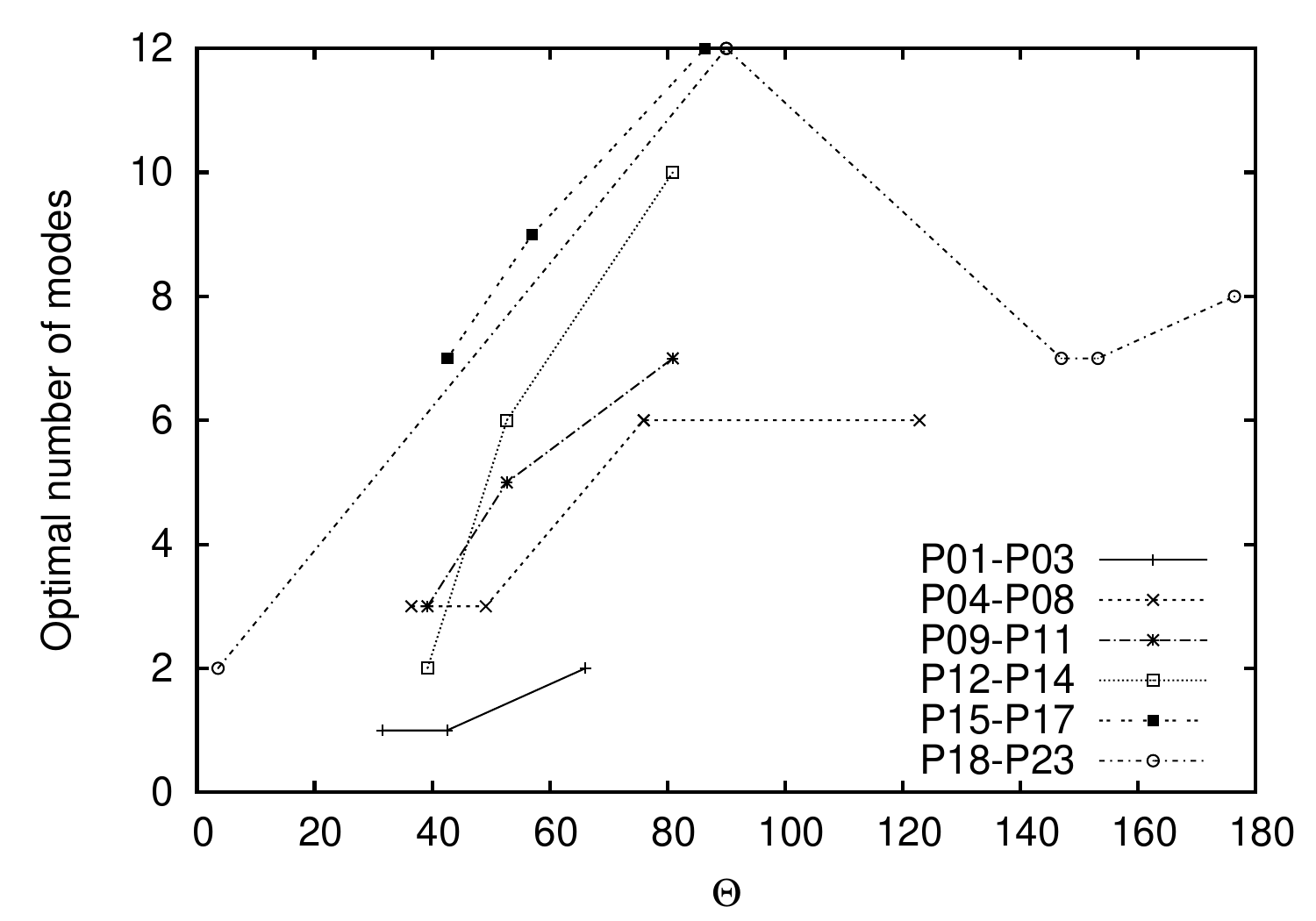}
}
\caption{  Optimal number of modes needed for full coverage versus $\Theta =   \mathrm{acos}{\left(\vec L \cdot \vec S / (L\,S)\right)}$ for
           all precessing simulations in the P-series.
}
\label{fig:preclines2}
\end{figure}

To quantify the influence of precession, Figure~\ref{fig:preclines2} shows the total number of modes
needed for full sky coverage in the P-series organized by mass ratio value $q$ and spin $\chi$. That is, lines join cases with the same mass ratio and spin.
The number of modes are given as a function of the angle $\Theta$ where 
$\cos{\Theta} =   (\vec L \cdot \vec S) / (L\,S)$.
Here $\vec L$ is the orbital angular momentum  and $\vec S = \vec S_1+\vec S_2$ the total spin at the beginning of the simulations.
Between $0^\circ \le \Theta \le 90^\circ$, the larger the angle $\Theta$ the higher the observed precession in the system. As mentioned before, a quadrupolar template is not
able alone to cover effectively the sky of orientations for these precessing binaries; consequently, it is not surprising to find in Figure~\ref{fig:preclines2} that the number of modes needed for full sky coverage grows monotonically in the interval $0^\circ \le \Theta \le 90^\circ$.
One can also see that in the same interval the lines are ordered from top-to-bottom with the mass ratio $q$, larger at the top and lower at the bottom. 
Therefore, for a given $\Theta$ angle value, the number of modes needed also increases monotonically with the mass ratio $q$. 
For angles  $\Theta \ge 90^\circ$, we only have a few data points and are not able to draw definitely conclusions. 

To check the robustness of our algorithm in finding hierarchies, we selected two $q=4$ binaries, one non-precessing  (Q06) and the other with one of the highest precession
(P23). We ran these two cases at different resolutions:  ($M/160,M/180,M/200$) for Q06 and ($M/140,M/160,M/180$) for P23. In addition, we constructed the \gw{s} from the simulations at three different extraction radii ($r=60,75,90M$).  
Our greedy algorithm was able to follow identical steps and find the same hierarchy for the Q06 case.
For the precessing case P23, there were small differences in the steps taken for the lowest resolution and smallest extraction radii case, but the
resulting hierarchy remained the same. 

We also experimented with steps that add and remove $(\ell,|m|)$ and $(\ell,-|m|)$ modes simultaneously. 
This reduced the number of total possible steps
from 31 to 17 after the initial seed of $(2,\pm2)$. The net effect was to shrink or eliminate Stage 2. 
We also found that 
there were very few cases where the resulting hierarchy contained fewer modes than 
the single mode case.

\subsection{Maximizing over intrinsic parameters}
\label{subsec:bank_maximization}

Because the focus of our study is orientation effects, we considered only the situation in which signal and the corresponding templates have the same 
intrinsic parameters (masses, mass ratios, and spins). In practice,  LIGO/Virgo searches the data is matched maximizing over intrinsic parameters against a bank of templates
spanning a range of masses and mass ratios~\cite{Babak:2012zx}.  For
the purpose of detection, it is only important that some template in
the bank respond to the signal with sufficiently high \snr{}, not
whether the parameters of this template match those of the signal.
Indeed, it is known that non-spinning templates can detect spinning
signals at the cost of incorrectly measuring $\eta$.  See, for
example, \cite{Aylott:2009ya}.  It is therefore appropriate to ask
whether the loss in overlap incurred by using only the (2,2) mode can
be compensated for by maximizing over the mass and mass-ratio.

To test this we again consider the set of orientations of a complete
waveform, represented as the sky centered at the source.  We take as
templates effective one body waveforms
(\texttt{EOBNRv2}~\cite{PhysRevD.84.124052})
as implemented in the  LIGO Algorithm
Library \texttt{lalsuite}~\cite{lalsuite}, as is done in current high-mass searches.
For each orientation we maximize the overlap over total mass $M$ and symmetrized mass-ratio $\eta$ using 
differential evolution~\cite{Storn10.1023}, a robust hill-climbing
algorithm.  This procedure overestimates the ability of the bank to
recover the signals, as in practice the bank is comprised of a
discreet set of templates arranged so as to lose no more than 3\% of
SNR~\cite{Babak:2012zx}.

\begin{figure}[tb]
\centering
\vbox{
\includegraphics[width=.95\linewidth]{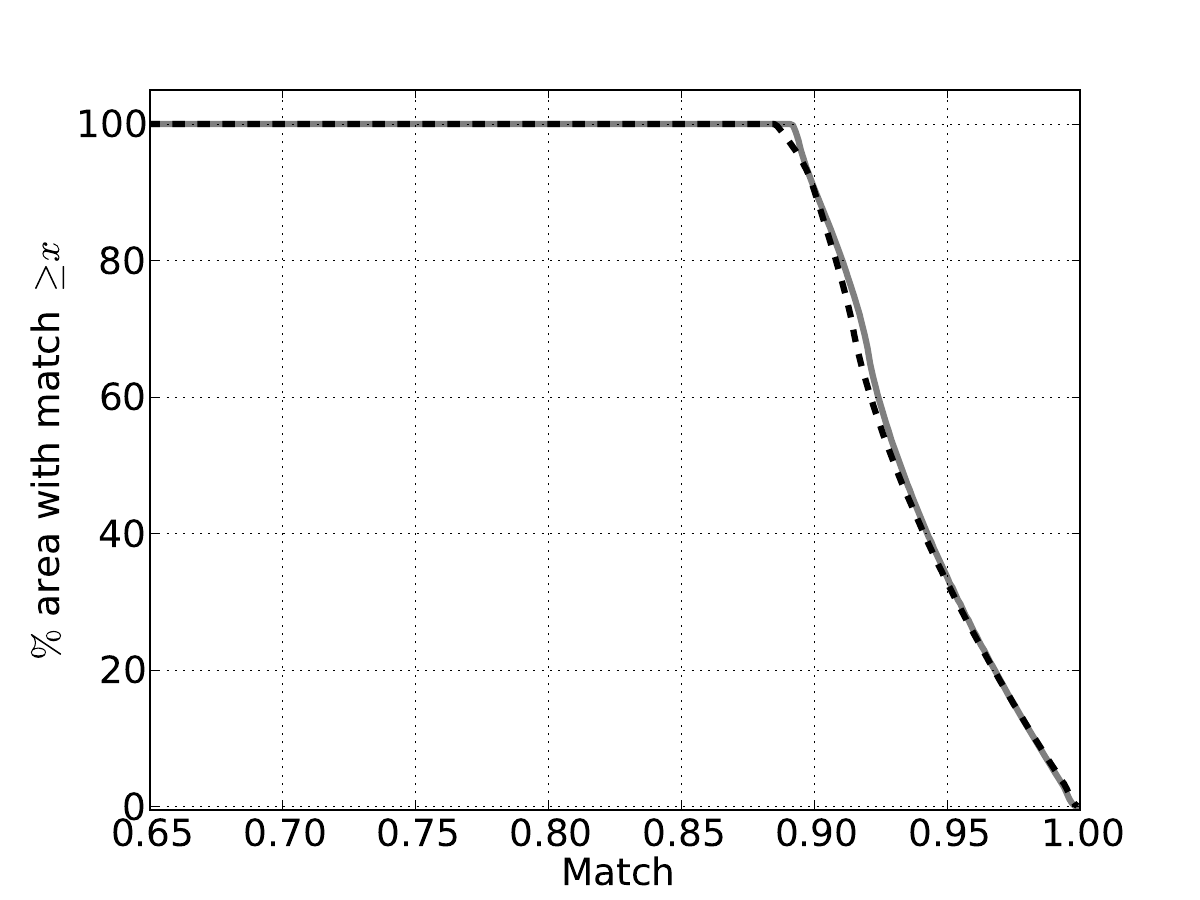}\\
\includegraphics[width=.95\linewidth]{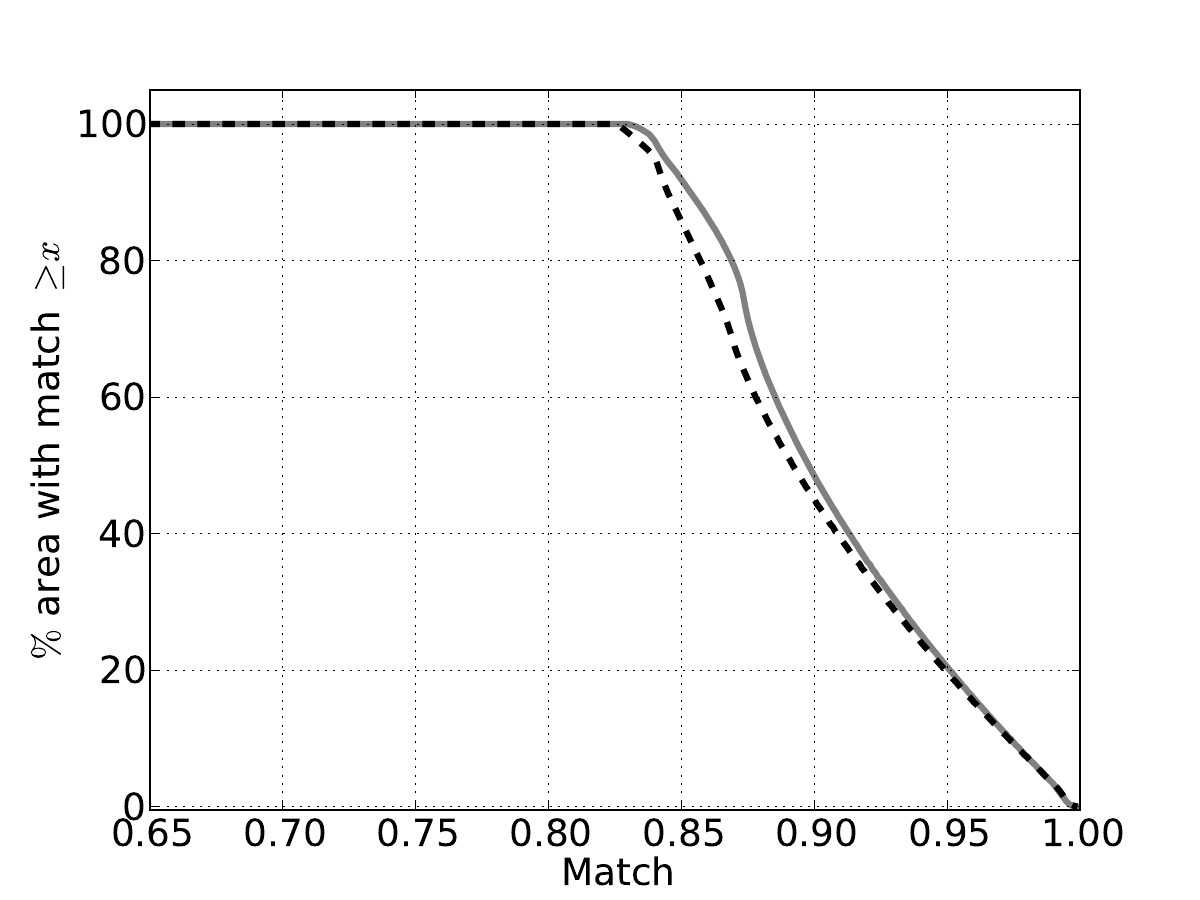}
}
\caption{Fraction of the sky area covered by EOB templates maximized 
over mass and mass ratio, as a function of match
for nonspinning $q=4$ (top) and $q=10$ (bottom) \bbh{} systems.
Dashed black lines show values obtained matching the NR signal to
itself, with no maximization.  Solid grey lines show values obtained
using \texttt{EOBNRv2} templates, maximized over mass and mass ratio.
See text for
details.  In particular, if it is desired to lose no more than 3\% 
of overlap then maximization over these parameters adds essentially
no area.
}
\label{fig:bank_maximization}
\end{figure}

The results for $q=4$ and $q=10$ waveforms are shown in
Figure~\ref{fig:bank_maximization}, presented as the fraction of the
sky covered at or above every overlap value.  The conclusion is that
maximizing over these parameters does not recover the loss in match
(hence \snr{}, distance, volume and event rate) incurred by neglecting
higher modes.  This result is not surprising in light of
Figure~\ref{fig:recomposedQ10}, as the recomposed waveform exhibits
features not present in any single mode.

\section{Conclusions}

As the NR community explores the complete, generic parameter space of
BBH spacetimes, non-dominant modes play an increasingly important role
in detecting and characterizing potential gravitational wave signals.
In this paper, we introduced a hierarchical, greedy method to identify
the modes necessary for full sky coverage.  We define sky coverage as
the percent of sky in a source-centric frame that a template will match
with nature up to a mismatch threshold, here taken as 0.03. 

The loss of sky coverage is most noticeable in highly unequal and
precessing BBH systems.  As long as $q \le 1.5$, the (2,2) mode will
effectively cover the entire sky for spinning and mildly precessing 
systems. For non-spinning systems, as the mass ratio increases to 
$q \ge 7$, the inclusion of four modes is necessary.  In the absence 
of precession, a larger spin magnitude increases the energy of the 
system and fewer modes are necessary.  As precession becomes more 
dramatic, as measured
by $\Theta$, the range of sky covered decreases to below 60-90\% and the
number of modes necessary to achieve full coverage increases to 12 for
the many of the cases studied here.  For non-precessing systems, the
four important modes are also the four most energetic: (2,2), (3,3),
(4,4), and (2,1).  For all precessing systems covered here, all the
$\ell=2$ and $\ell=3$ modes are sufficient for covering all possible
orientations.

Although we have focused on binaries with total mass of $100\,M_\odot$, 
our methodology is directly applicable to less massive systems, where
templates are constructed from hybrid waveforms that combine post-Newtonian and NR
results into one, long waveform.  In broad terms, our conclusions regarding the influence of higher modes are likely to carry over 
to lighter binaries, see for instance Figure 1 and related work
in~\cite{Brown:2012nn}, and thus have an important impact for the enterprise of building template
models. Our method could potentially give clues where effort in analytic modeling, NR
simulations, and hybridization techniques will yield the most
benefit.  In particular, it should be possible to estimate the loss in sky coverage 
due to missing information resulting from using a subset of modes.

Our results suggest that a substantial loss of overlap over a
significant fraction of orientations could result without the inclusion of higher
order modes in templates.  Furthermore, this loss can not be
alleviated by allowing the template intrinsic parameters (masses and
mass ratios) to vary.  
However, it remains to be investigated whether the inclusion higher modes improves
the efficiency of real searches.  A template bank
responds not only to \gw{} signals, but also to noise in the detector,
creating a population of false alarm triggers in the search pipelines.
As a consequence, a true signal must stand out sufficiently above
those triggers in order to be claimed as a detection.  If one were to
utilize higher-mode information, it would become necessary to increase
the dimensionality of the template bank by two to include the
orientation parameters $\theta $ and $\phi$.  There is a potential
tension in doing so.  Although the additional templates will assist in
responding better to the presence of signals, the increased number of
templates will unfortunately also elevate the population of false
alarm events thus impairing detection ability.  In order to maximize
the template response and minimize the emergence of false alarm
triggers, one would have to adjust various aspects in the current
search methodologies.  For example, currently two distinct searches
are performed, one covering the mass region $M \le 25 M_\odot$ and
another spanning $25 M_\odot < M \le 100 M_\odot$.  This is done in
part because the background populations generated by templates in
these two regions differ significantly.  It is possible that
higher-mode information could best be employed by 
further subdividing by mass ratio.  

There is a potential follow-up to this work that avoids the
difficulties inherent in constructing a full higher-mode search.  A
population of NR signals randomly distributed in distance,
orientation, and Earth-centered sky location could be injected into
real or realistic detector noise.  By running the full search over the
resulting data it would be possible to determine the detection
efficiency as a function of distance.  By repeating this process
twice, once injecting only the dominant mode and once injecting
complete waveforms, it would be possible to experimentally determine
the degree to which higher modes in real signals reduce the search
efficiency.  The NINJA-2
project~\cite{Ajith:2012tt}, which is currently ongoing, provides one
possible context in which such a study could be performed.  This study
would be computationally expensive and would require expertise and
infrastructure currently only available in the LIGO/Virgo
collaboration.  We suggest that a major result of the present work is
that such a follow-up study is justified.

\section{Acknowledgments:} 
\label{sec:acknowledgements}
Work supported by NSF grants 1205864, 1212433, 0903973,  
0941417 and 0955825. Computations at XSEDE  TG-PHY120016, TG-PHY090030 and the Cygnus cluster at Georgia Tech.

\bibliography{nr}

\end{document}